\documentclass[titlepage,twoside]{article}
\usepackage[american]{babel}
\usepackage[utf8]{inputenc}
\usepackage[T1]{fontenc}
\usepackage[backend=biber,useprefix=true,style=ieee,url=true,dashed=true,doi=false,eprint=false]{biblatex}

\usepackage{csquotes}

\usepackage{subcaption}

\DeclareLanguageMapping{american}{american-apa}

\DeclareUnicodeCharacter{0301}{\'{e}} 

\bibliography{references-NRXR-ID}

\usepackage[breaklinks, colorlinks=true, allcolors=black]{hyperref}
\usepackage{amsmath}
\usepackage[rightcaption]{sidecap}
\usepackage{caption}
\usepackage{sidecap}
\usepackage{placeins}
\usepackage{float}
\usepackage{MnSymbol} 
\usepackage{enumerate} 
\usepackage[table]{xcolor} 
\usepackage{multirow} 
\usepackage{adjustbox} 
\usepackage{graphicx} 
\usepackage{longtable} 
\usepackage{booktabs} 

\usepackage{ragged2e}

\usepackage{verbatim}

\graphicspath{ {./figures/} }

\usepackage{textcomp}
\usepackage{gensymb}

\usepackage{indentfirst}

\makeatletter
\renewcommand*{\@fnsymbol}[1]{\ensuremath{\ifcase#1\or *\or \dagger\or \ddagger\or
   \mathsection\or \mathparagraph\or \|\or **\or \dagger\dagger
   \or \ddagger\ddagger \else\@ctrerr\fi}}
\makeatother

\usepackage{fancyhdr,lipsum}
\begin{document}

\title{NRXR-ID: Two-Factor Authentication (2FA) in VR Using Near-Range Extended Reality and Smartphones}


\author{Aiur Nanzatov$^1$ \and Lourdes Pe\~na-Castillo$^1{^,{^2}}$\and Oscar Meruvia-Pastor$^1{^,*}$\thanks{*Corresponding author: Oscar *at* MUN *dot* ca}}

\date{
	$^1$Department of Computer Science, Faculty of Science, Memorial University of Newfoundland, Canada \\ 
	$^2$Department of Biology, Faculty of Science, Memorial University of Newfoundland, Canada%
}

\maketitle

\abstract{
Two-factor authentication (2FA) has become widely adopted as an efficient and secure way to validate someone's identity online. Two-factor authentication is difficult in virtual reality (VR) because users are usually wearing a head-mounted display (HMD) which does not allow them to see their real-world surroundings. We present NRXR-ID, a technique to implement two-factor authentication while using extended reality systems and smartphones. The proposed method allows users to complete an authentication challenge using their smartphones without removing their HMD. We performed a user study where we explored four types of challenges for users, including a novel checkers-style challenge. Users responded to these challenges under three different configurations, including a technique that uses the smartphone to support gaze-based selection without the use of VR controllers. A 4X3 within-subjects design allowed us to study all the variations proposed. We collected performance metrics and performed user experience questionnaires to collect subjective impressions from 30 participants. Results suggest that the checkers-style visual matching challenge was the most appropriate option, followed by entering a digital PIN challenge submitted via the smartphone and answered within the VR environment. \\\\
\textbf{Keywords:} 2FA, Smartphones and HMDs, Phone in VR, Virtual Reality and Human-Computer Interaction, Information Security, VR Interaction Techniques,   Authentication in VR, Two-Step Authentication

}

\begin{figure*}
  \includegraphics[width=\textwidth]{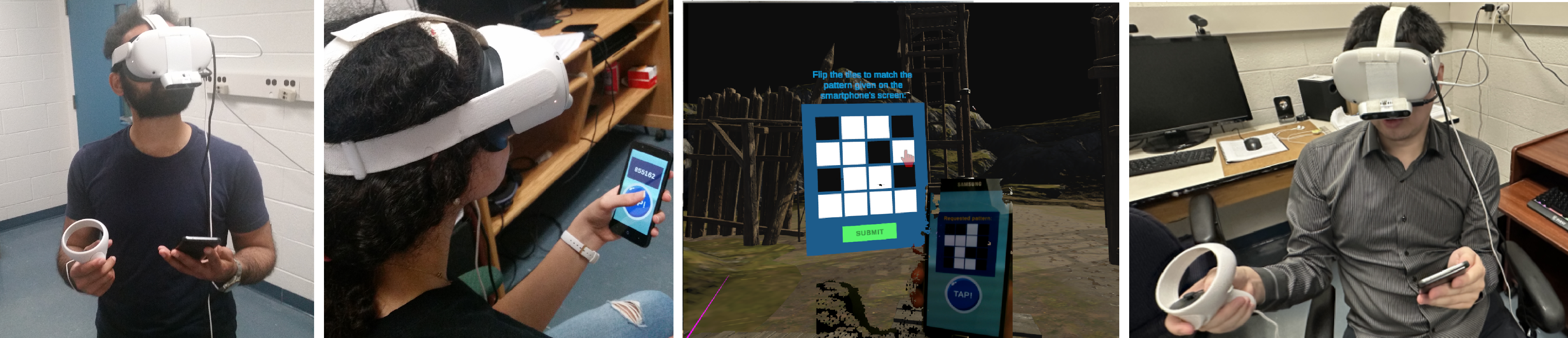}
  \caption{Showcase of different conditions, left to right: with the smartphone and the VR controller; with the smartphone only; VR environment view; operating the scene with the VR controller.}
  \label{fig:teaser}
\end{figure*}


\maketitle

\section{Introduction}
This research explores the  design space of multi-device authentication methods that combine smartphones and VR headsets, with a focus on supporting two-factor authentication (2FA) for VR applications. 

Authentication methods have evolved significantly, driven by the ongoing battle between security developers and cyber attackers~\cite{LandwehrCAPTCHA, Al-Hasan, Flores, Al-Hadadi, Henrysson, authenticationMechanisms}. As hackers and automated systems become more sophisticaded, innovative protection mechanisms continue to evolve as well~\cite{MylonasSPSecurity, SPisSafe, StackPole, Mashkina}. This dynamic highlights the need for security solutions that are both robust and user-friendly, balancing security and strong protection of privacy with seamless user experiences.


While there are continuously emerging security risks that are particular to XR systems, such as vulnerabilities in the VR devices themselves~\cite{Sha:2025}, gaze exploitation~\cite{Wang:2024}, 3D disguise~\cite{Fujita:2023,Lebeck} and de-anonymization attacks~\cite{Sabra:2024}, many security risks associated with the use of extended reality systems are are also present in non-VR systems~\cite{Guzman}, including identity theft and impersonation, data privacy breaches, and man-in-the-middle, phishing, and social engineering attacks~\cite{Peter:2024}. 

As technology progresses, safeguarding user credentials requires increasingly robust  strategies~\cite{SecureProtocol, dastgerdy2024virtualrealityaugmentedreality}. In this context, two-factor authentication (2FA) has become a widely used method for secure identity validation, providing enhanced protection by requiring the user to possess two separate devices for authentication. 
Our focus on two-factor authentication in VR is justified by its widespread adoption and effectiveness in improving security. Traditional single password-based systems remain vulnerable to attacks such as phishing, advances in artificial intelligence malware, and data breaches, making them insufficient on their own for safeguarding sensitive data~\cite{Jubur:2025}. There are two stages in two-factor authentication. In the first stage, a user gains access to the system, for example by providing a password. In many cases, that is considered enough, for example when a user regularly accesses an email account from the same web browser of a particular computer. However, in some cases, a system flags a situation that requires a confirmation of identity (for example, when the user tries to access their email from a new computer), and this is where the the user is required to check the second device to find a challenge that they must provide to confirm their identity. 

The combination of passwords and two-factor authentication has emerged as a key solution for safeguarding security and  privacy in non-VR systems~\cite{Jubur:2025}.  In fact, an increased use of 2FA across various industries including technology, banks, social
media, and health organizations has been observed recently~\cite{Bhanderi:2023}. In particular, one-time personal identification numbers (PINs) sent using email and SMS have become the most commonly used two-factor authentication method in the financial sector~\cite{Bhanderi:2023}. In most cases, the second device used for authentication is a smartphone, where users  receive a text-message to the phone number associated with their online account. The text message typically contains a numeric code (PIN) that they need to enter in the primary device where the need for authentication was generated in the first place. In other cases, users receive the code in an alternative email account and are requested to check their email account to find out the code. There are many situations where the need to perform two-factor authentication may occur within a VR environment. For example, a user who needs to purchase a valuable asset in a VR game could be asked to authenticate during the game. Similarly, a health-care provider in VR could be asked to confirm their credentials to access a new set of sensitive  medical records in a tele-health scenario. In a third scenario, a VR user using a web browser in the Metaverse may be asked to authenticate their identity to access their email account or to confirm a commercial transaction flagged as suspicious by their financial institution.

In Virtual Reality (VR), performing two-factor authentication  becomes more challenging than usual because users typically cannot see the devices in their real-world surroundings while wearing a head-mounted display (HMD). Instead, a user would typically be forced to lift their HMD to read the code or access their email account on a smartphone or nearby desktop computer, or they would need to step out of the play area to be able to see their smartphone using pass-through cameras mounted in the HMD.  This could potentially disrupt their VR experience or their activities within the VR application where the need for authentication may have originated.

\subsection{Near-Range Extended Reality for 2FA}


Near-Range Extended Reality (NRXR) refers to the type of immersive VR experiences that integrate real-world elements in close proximity to the user~\cite{McGill:2015,Budhiraja:2015}. The need to become aware of the real-world surroundings and  the actions of others in close proximity to those within VR has existed since VR's origins~\cite{Foerster, NeverBlindVR, Hartmann:2019, McGill:2015, Budhiraja:2015,Kanamori:2018}. 
While substantial work has been done in the area of integrating personal devices, such as tablets~\cite{Surale:2019}, smartphones~\cite{ZhuGrossman:2020,Henrysson, Taejin2}, smart watches~\cite{Siddhpuria2018}, keyboards~\cite{McGill:2015}, and desktop workspaces~\cite{Wentzel:2024}, the specific topic of identity authentication using such devices in VR is much less explored.
To facilitate user authentication tasks without leaving the VR context, we propose the use of Near-Range Extended Reality  and suggest how it  can be used to support two-factor authentication. We refer to the use of NRXR for two-factor  authentication as NRXR-ID. NRXR-ID allows users to complete authentication using their smartphones without removing their HMDs. NRXR-ID prioritizes awareness of objects in close proximity of the user because in the context of authentication there is no need to become aware of elements from the real world that might be far from the user. In prior work, McGill et al~\cite{McGill:2015} found that selectively presenting some elements from reality as users engage with VR allows for optimal performance and maintains the users' sense of presence. By using NRXR, users can access their smartphones, nearby laptops, or personal computers while staying aware of the VR environment. This is particularly important when users need information from the real world and the VR environment to complete the authentication procedure.

For this project, NRXR was implemented as a technique for VR systems where a depth-sensing camera mounted on the HMD allows users to see nearby objects (i.e., those 10-120 centimeters away from the camera) within the virtual environment, with the main goal being  that users are able to see the smartphone from within the VR. This implementation creates a unique form of Mixed Reality (MR) experience where physical objects that are close to the user become visible in the virtual world in a more selective way~\cite{EnhancingReality, Tecchia:2014, McGill:2015}, differing from Meta's Passthrough or the Vision Pro's crown dial in the way in which real-world imagery is blended into the VR experience.  We implemented Near-Range eXtended Reality with an external Depth-sensing camera because the SDK for the Meta Quest 2 did not provide access to depth field video stream or the passthrough cameras, which in turn prevented us from getting a depth video stream to filter for nearby objects in front of the user. For widespread deployment, developers with access to the built-in cameras and/or depth sensors in newer Mixed Reality HMD’s should be able to implement NRXR without the need to mount an external depth-sensing camera, as long as those HMD's can convey a depth field as seen from the user's perspective.

With the latest generation of VR HMDs like the Meta Quest 3, Varjo XR4, and Apple Vision Pro, which support Mixed and Augmented Reality (MR/AR), the ability to combine the real and virtual worlds to enhance the VR user experience has become much more convenient. Recent MR/AR solutions include the “Passthrough” mode in devices like the Meta Quest and the Apple Vision Pro. In contrast to prior implementations of NRXR~\cite{McGill:2015,Tecchia:2014,Budhiraja:2015}, these solutions do not adjust the amount of the real world that is shown to the users as a function of the distance between objects in the real world and the user. For example, the Passthrough mode in the Meta Quest allows users to see their physical surroundings through built-in cameras by default when they step outside the virtual boundaries, or by means of Mixed Reality windows to the user surroundings using passthrough components in Unity. However, depending on the use case scenario, asking users to step out of the VR could completely disrupt the VR experience. As an alternative passthrough design, a physical ``Crown'' dial in the Apple Vision Pro lets users manually adjust the passthrough region in their field of view. In this case, the VR imagery can be extended from the center of projection to the sides, as if holding a curtain that can be manually opened of closed in front of the viewer, while a passthrough view of the real world shows up in the peripheral regions of the user's field of view. However, this manual adjustment is independent of events happening outside of the VR world and is not necessarily meant to allow nearby objects to become visible or increase security, but is rather meant to allow objects in the peripheral vision region to become visible in lieu of a VR background. Despite these differences in passthrough designs, it can be argued that the newer, higher-end HMDs are well positioned to support 2FA by offering a more convenient setup (including an HMD with no camera attached, better image and cameras resolution, and hand tracking). In fact, the Apple Vision Pro already allows users to see their smartphone while wearing the HMD, and we discuss its viability for 2FA, as well as that of the Meta Quest 3 and Meta Quest Pro in the Future Work section. These latest advances in HMD technologies allows us to focus on questions related to the design choices available to make an XR implementation of two-factor authentication that is useful, convenient, and preferred by users, as explored in the next sections.

\subsection{Research Questions}
 In this section, we describe the research questions that will guide our exploration on how two-factor authentication can be supported in VR, Mixed Reality (MR), and Augmented Reality (AR) using NRXR. 

Going back to the topic  of authentication, it is crucial to design authentication methods that offer both robust security and practical convenience~\cite{Biddle}. This research seeks to answer questions about the best way to help VR users perform two-factor authentication without needing to remove their HMD or lose awareness of what is happening in the VR environment.
The main research questions we address in this work are:
\begin{itemize}
\item RQ1: Is it possible to use NRXR to support people in the process of confirming their identity via two-factor authentication?
\item RQ2: What type of authentication challenges and modalities are most suitable to implement a user-friendly form of two-factor authentication in the XR context?
\item RQ3: When implementing two-factor authentication using NRXR, what is most effective for users: to present a challenge using the smartphone and have the challenge answered within the VR, or the other way around?
\item RQ4: What are users’ experiences, impressions and preferences on two-factor authentication when using NRXR?
\end{itemize}

To answer these research questions, we have evaluated several forms of  NRXR-ID with a user study described in the Methodology.

\begin{figure*}[h]
  \includegraphics[width=0.99\textwidth]{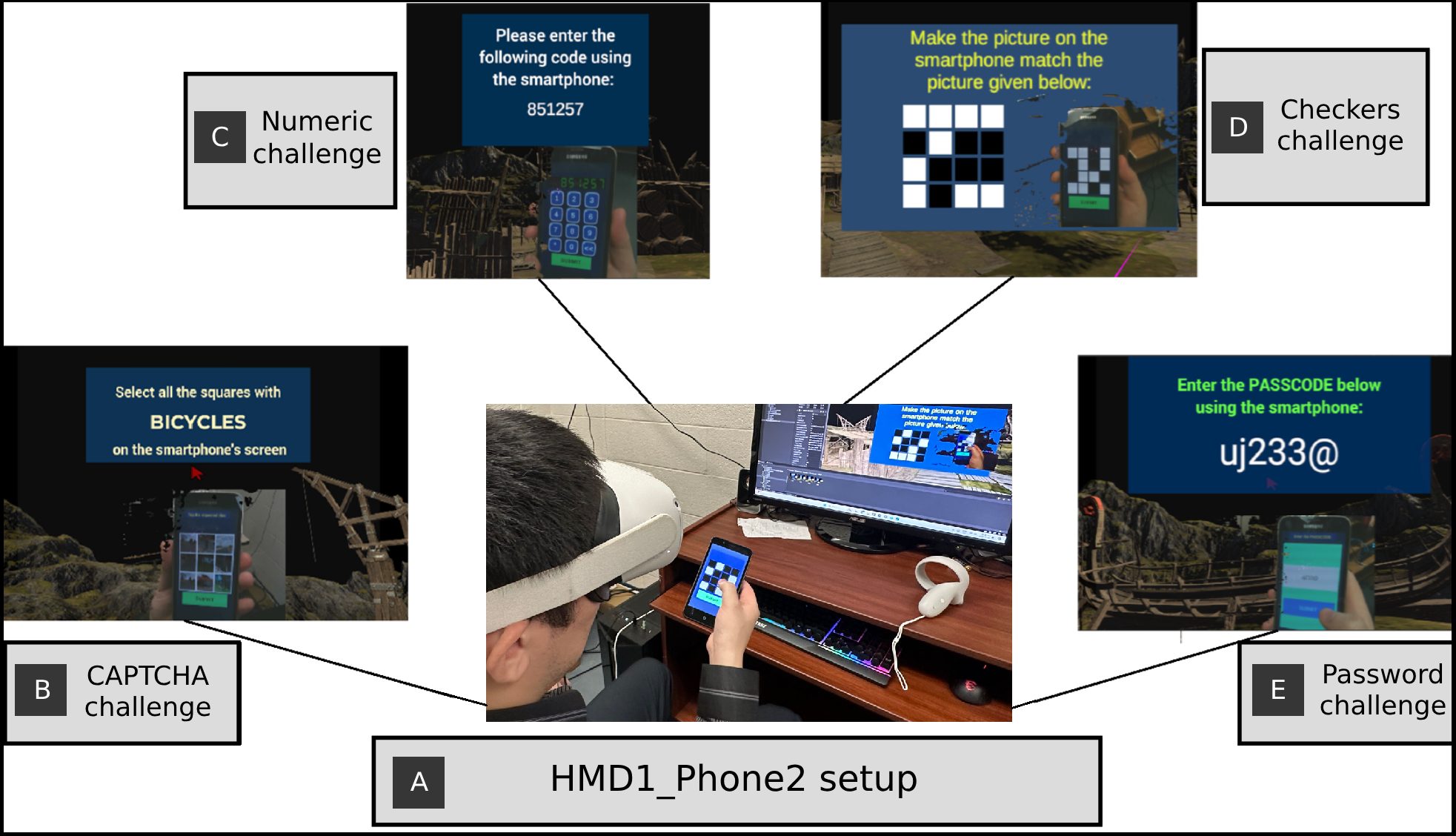}
  \caption{Overview of the authentication methods of condition 1 (HMD1\_Phone2): A) Illustrates the experimental setup with a user holding the smartphone and completing the challenges; B,C,D,E) Sequence of tasks provided for the participants to solve using the smartphone app, correspondingly: CAPTCHA challenge, Numeric challenge, Checkers challenge, Password challenge.}
  \label{fig:challenges_condition1}
  
\end{figure*}

\section{Related Work}
There are multiple methods for authentication tailored for AR/XR. Behavioral biometrics, based on users’ unique behavioral patterns, offer a non-intrusive authentication method. For instance, hand tracking in AR/VR environments can identify users based on their finger movements and gestures. Liebers et al.~\cite{Liebers} demonstrates the effectiveness of hand tracking for implicit user identification in immersive environments.

Other novel VR authentication methods include: 1) \textit{Gesture-Based Authentication}, where users authenticate through personalized gestures like high fives or fist bumps, maintaining immersion and improving security~\cite{VirtualAgents}; 2) \textit{Direction-Based Authentication} (DBA), where users navigate virtual environments and select directions to form a password, balancing memorability, efficiency, and security~\cite{DBA}; 3) \textit{SPHinX}, a method where users paint or trace patterns on a 3D object to enhance security and reduce risks like shoulder surfing~\cite{SPHINX}. These methods aim to integrate authentication seamlessly into the VR experience while addressing security concerns, and could be used to gain initial access to a system, but are not suitable to support multi-device authentication.

Biometric authentication provides a viable alternative in VR, where traditional methods like PINs and passwords are difficult to implement.  Biometric authentication methods include retinal scanning~\cite{Vora:2012}, iris scanning~\cite{Daugman:2014}, and skull bone conductivity~\cite{Schneegass}. A study by Heruatmadja et al.\cite{Biometry} reviews biometric techniques for identification in VR and highlights the accuracy of finger vein and hand movement biometrics, using machine learning methods like k-Nearest Neighbors (k-NN) and Support Vector Machines (SVM). These techniques would first need to be made available in HMD's and could then be used for the first stage of the two-factor authentication process to gain initial access to a system, whereas smartphones could be used for the second step of authentication, i.e., confirming the user's identity.

Recent research has explored the use of traditional password-based authentication in VR and AR environments~\cite{PicturePasswords}. Entering passwords in these settings can be cumbersome and negatively impact the user experience~\cite{VirtualAgents, authImmersiveAnastasaki}, highlighting the need for more intuitive solutions. Alternatives like virtual or touch-sensitive physical keyboards are promising for text entry in VR environments~\cite{Menzner}. While physical keyboards are effective in VR, they require external camera-based tracking systems~\cite{Menzner}. Touch-sensitive keyboards, which track fingertip movement directly on the keyboard’s surface, offer a potentially intuitive and accessible option for password entry in VR.

Decentralized technologies and self-sovereign identity (SSI) offer a promising alternative for enhancing VR authentication. SSI mitigates vulnerabilities in traditional methods, such as predictable passwords and biometric theft, by providing a decentralized framework that gives users control over their identity. It incorporates memory-based authentication, where users recall and create scenes stored on the InterPlanetary File System (IPFS) and blockchain, reducing risks from centralized data storage~\cite{SSI,Mchale:2023}. However, its success depends on overcoming challenges like memory recall, technical integration, scalability, and user acceptance.

While the solutions above have explored particular forms of single  authentication in XR, this paper focuses on the confirmation of identity stage of 2FA. This research is among the first empirical studies to compare different types of challenges involving multi-device authentication while exploring different device configurations.

\subsection{Using smartphones in VR}

Several research projects have explored integrating smartphones~\cite{Desai2017, Alaee2018, Mohr2019, Hattori2020, Zhang2021, Bai2021, Unlu2021} and other input devices, such as handheld controllers~\cite{Hincapie-Ramos2015, Young2017} and smartwatches~\cite{Kharlamov2016, Kim2016, Hirzle2018, Park2019}, in virtual, augmented, and mixed realities. For instance,~\cite{Pietroszek2014} utilizes smartphones as input devices for interacting with displays without expensive tracking devices, and~\cite{Aseeri2013} proposes using mobile devices for various interactions in VR. Other handheld devices like touchpads have also been explored for VR interactions~\cite{Budhiraja2013}.

In~\cite{SVRP:24}, smartphones are used for selection tasks and teleport-based navigation in VR, showing they can be comparable to VR controllers. In~\cite{Bai2021}, the concept of AV is used to access smartphones, emphasizing the importance of including users' hands and realistic skin tones to enhance interaction. NRXR incorporates these important features. Zhu et al.~\cite{PhoneInVR2024} explore how smartphones should be spatially anchored in VR, suggesting that physically holding smartphones and using direct touch improves accuracy and speed. This concept aligns with how NRXR provides users access to their smartphones in VR.

\section{Methodology}

To assess the limitations, advantages, and drawbacks of our proposed system, we considered various ways two-factor authentication can be implemented. While the number of potential implementations is large, we have focused on four types of challenges for this initial exploration. A user study is conducted to compare these four strategies in terms of efficiency and user satisfaction.

\begin{figure}
    \centering
        \includegraphics[width=\columnwidth]{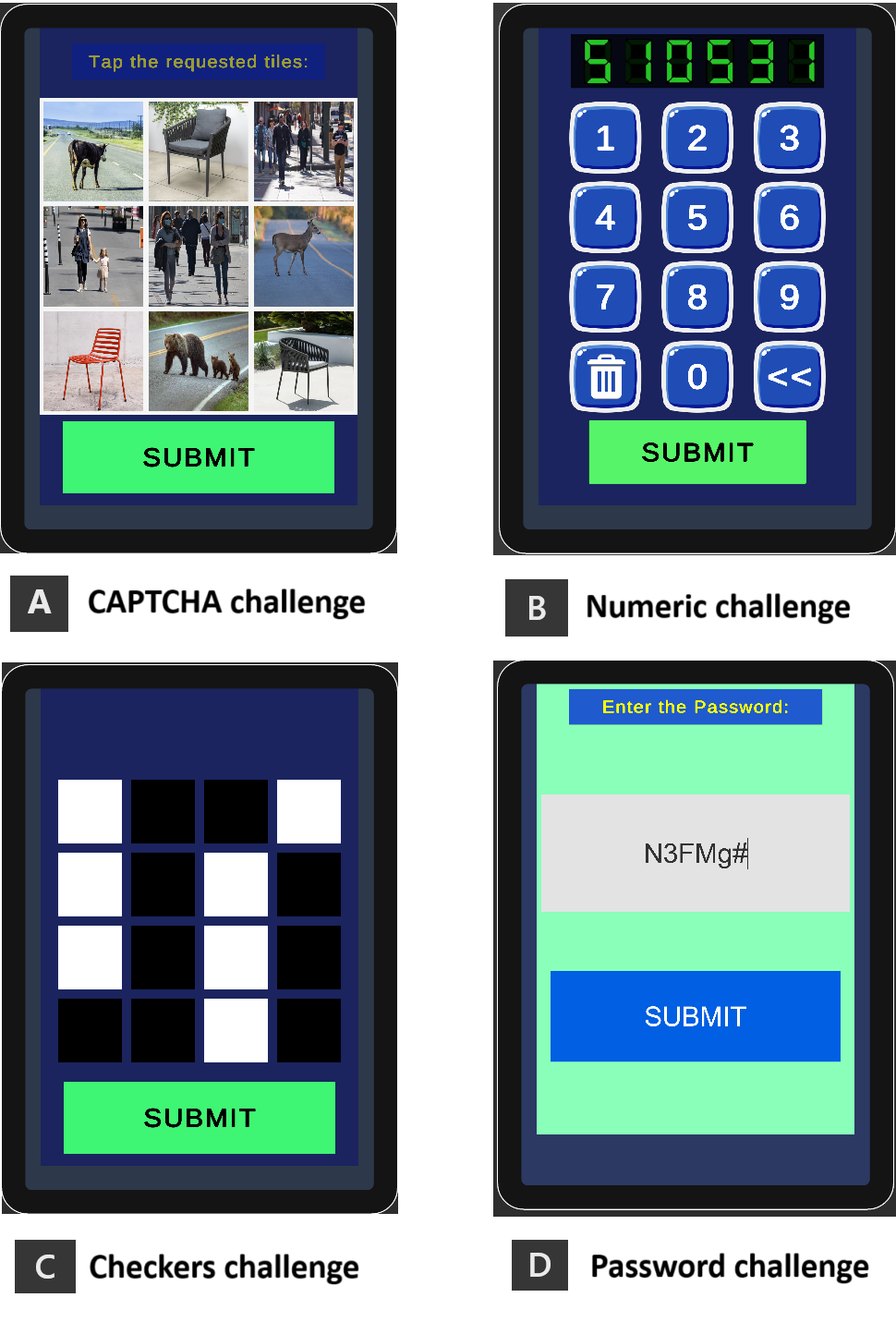} 
        \caption{Overview of challenges: A) CAPTCHA-style - selection of tiles corresponding to a request sent via HMD. B) Numeric - enter a six-digit code. C) Checkers - visual matching of two checkered grids. D) Six-character alphanumeric password.}
        \label{fig:SmartphoneApp_condition1}        
\end{figure}

Given the variety of possible approaches, we have chosen to focus on four specific authentication methods: solving a CAPTCHA challenge (a well-established verification method), entering a numeric PIN code (the most common form of two-factor authentication), checkers-style matching (a visual matching challenge suitable for graphical interfaces), and providing an alphanumeric password using a virtual keyboard. Figure~ \ref{fig:SmartphoneApp_condition1} illustrates these four forms of authentication, which will be described in detail in the next sections.

\subsection{CAPTCHA-style challenge}
In the CAPTCHA challenge, the user is presented with a 3x3 grid of images and must select three objects that match the description provided in the first step of the challenge~\cite{LandwehrCAPTCHA}. Only the challenge creator knows which items in the grid correspond to the correct answer. To implement this, the challenge creator sends a general description of the relevant icons to one device and the challenge grid to the second device used for authentication. For example, the user might be asked to select all three images containing animals on one device, while the second device displays a grid with animals and other content. There are three "themes" that can be fit into a 3x3 set of tiles and users need to select three tiles to solve a challenge. Thus, users must complete two rounds of CAPTCHA challenges, requiring at least six clicks in total to solve the task. 

\subsection{Numeric code challenge}
One of the most common forms of two-factor authentication involves sending a 6-digit numerical code to the user's smartphone~\cite{Jubur:2025}. Typically, the user receives this code via text message and then enters it on a second device. This method is widely used by banks, email providers, and other online services. In our implementation, we replicate this process in a custom smartphone app. The user receives the numerical code on one device and is then prompted to enter it on the second device.

\subsection{Checkers matching challenge}  
The checkers-style visual matching  challenge was devised as a form of visual authentication suitable for graphical interfaces. Its tiled arrangement resembles the CAPTCHA-style challenge, and was designed with visual simplicity in mind. The user receives a  4-row by 4-column (4x4) grid of checkered tiles in one device and receives a second 4x4 grid of checkered tiles on the second device. The task consists of flipping each tile between black and white states by tap or selection, until the tile arrangement shown in the second device matches the arrangement shown in the first device. There are only 6 differences in the tiles, so the user could solve the challenge by doing as little as 6 flips. Compared with  the images shown in the CAPTCHA-style challenge, which require recognition of the contents of each picture in the grid, the checkers type of challenge has the advantage that the tiles have the maximum contrast possible and the tiles are easily identifiable as being on or off. Finally, the tiles can  be easily encoded as a compact bit sequence, encrypted/decrypted, and converted to a visual challenge, or the other way around, once a user has submitted a response. 


\begin{figure*}
    \centering
        \includegraphics[width=\textwidth]{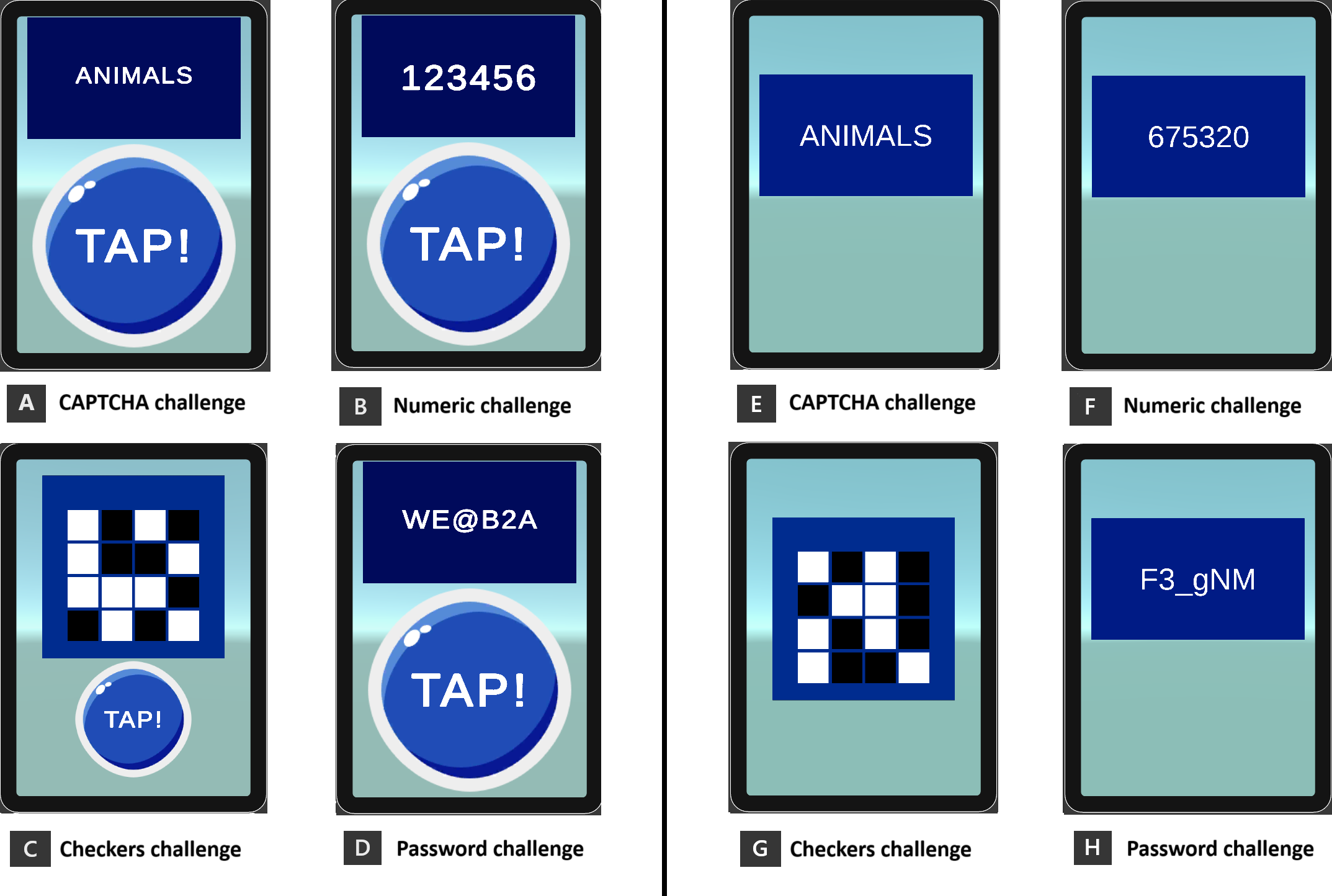} %
        \caption{Overview of the authentication methods used for step 1 during conditions 2 and 3 (Phone1\_SVRP2 and Phone1\_VRC2): A-D) Exemplifies  challenges presented on the smartphone when using Phone1\_SVRP2. The tap button allows users to indicate their selection within the VR using a gaze-based pointer~\cite{SVRP:24,Pathmanathan:2020}. E-H) Exemplifies  challenges presented on the smartphone when using Phone1\_VRC2, where participants use the VR controller's trigger button to indicate their selection within the VR.}
        \label{fig:Phone1challenges}
\end{figure*}

\subsection{Alphanumeric Password challenge}
Alphanumeric passwords are often used as the first line of authentication for users of online systems. Given the inherent vulnerabilities of password-only authentication, two-factor authentication has emerged as a crucial supplementary measure~\cite{Jubur:2025}. In this challenge we have kept some of the established rules for creating robust passwords: use of upper and lower case, and combination of numbers, letters and special characters. To keep this challenge complexity to a level comparable to the other authentication strategies, we added a restriction that the password will have maximum length of 6 characters, even though most passwords are usually a minimum of 8 characters. This keeps the number of required clicks from the users comparable to that of the other three challenges, but users still have to switch between the different character sets in the virtual keyboards to complete the challenge.

\subsection{Configuration possibilities for 2FA in XR}
In NRXR-ID, the smartphone can act as the device showing a 2FA challenge prompt, while the HMD can be used as the device to provide the solution to the challenge, or the other way around. Figure~\ref{fig:teaser} highlights some of the possibilities available for interaction. In this study, we explored three distinct configurations:  
\subsubsection{Condition 1: HMD1\_Phone2}
In this condition the HMD is used as the first device for the authentication challenge, i.e., the device that shows the user the challenge to be solved (step 1). The smartphone is used as the second device for the authentication, i.e., the device where the user enters and submits their solution to the challenge (step 2). For this condition a client app needs to be installed on the smartphone to capture and submit the participant's responses to the challenges posed in the VR application. Figure \ref{fig:challenges_condition1} illustrates how the challenges are communicated to the participant within the VR and how participants see the smartphone app while wearing the HMDs. Figure~\ref{fig:SmartphoneApp_condition1} shows the smartphone app screens corresponding to each type of challenge, and Figure~\ref{fig:exposure}C shows the smartphone's virtual keyboard used for answering the Password challenge. A user scenario for this modality is when a bank's client is accessing their bank account via a web browser shown in VR and needs to confirm their identity for a transaction that has been flagged by the bank, such as a stock market transaction. The bank then generates a token that the user can only enter using a bank-provided app that has been pre-installed on the client's smartphone. The user then grabs their phone, opens their bank's app and enters the token shown in the web browser using the app.

\subsubsection{Condition 2: Phone1\_SVRP2 }
\begin{figure}[htbp]
    \centering
    \includegraphics[width=\columnwidth]{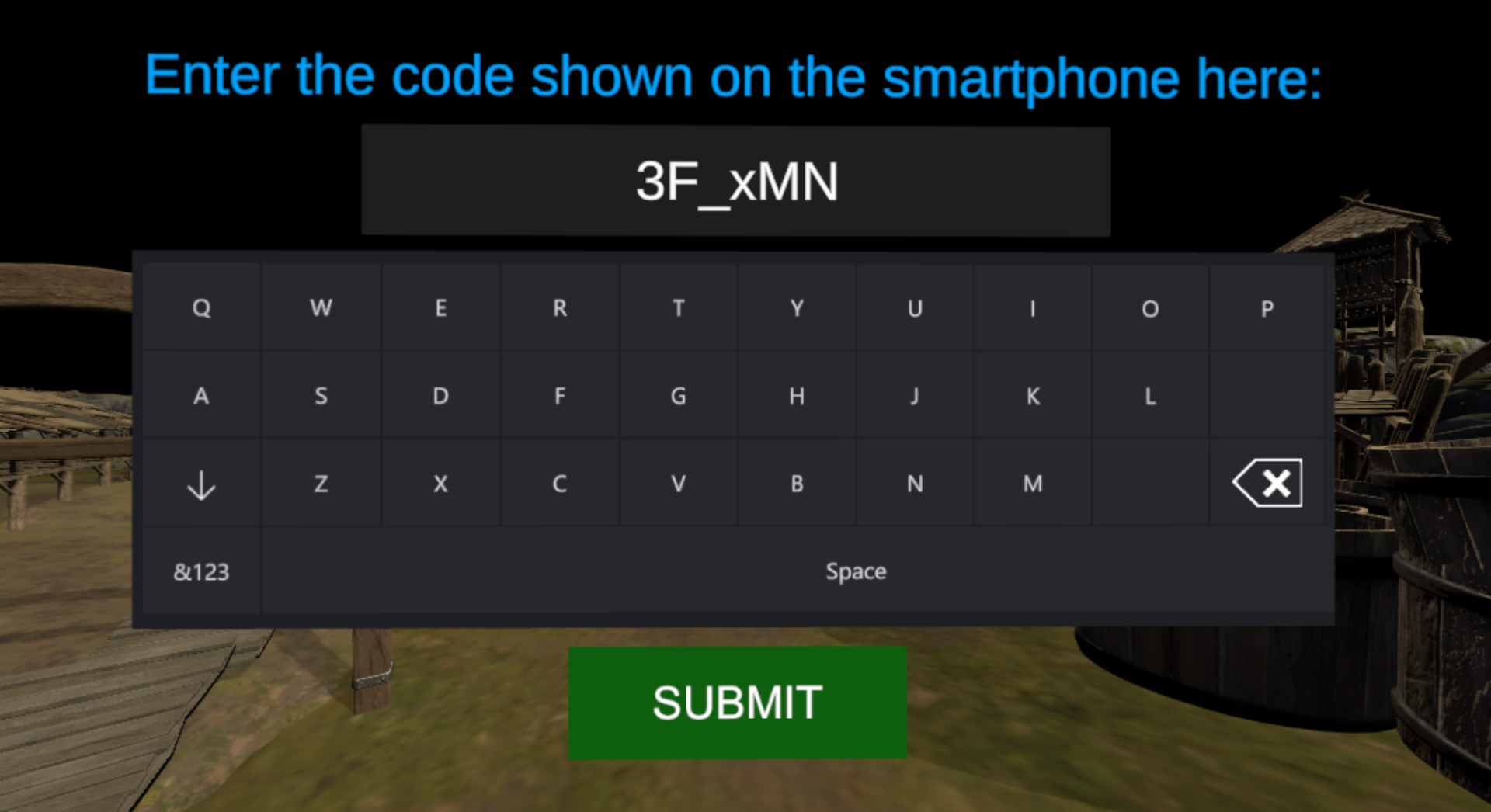}
    \caption{Virtual Keyboard in the VR environment. Utilized during Password challenge in conditions 2 (Phone1\_SVRP2) and 3 (Phone1\_VRC2) }
  \label{fig:keyboard}
\end{figure}

In this case the smartphone is used as the first device for authentication, i.e., the device that shows the user the challenge to be solved (step 1, Figure~\ref{fig:Phone1challenges}, left panel), and the HMD is used as the second device for the authentication, i.e., the device where the user enters and submits their solution to the challenge (step 2, Figure ~\ref{fig:challenges_condition2}). Since the challenges  can be solved within the VR environment with different input devices, users need a selection mechanism to either select tiles from grids, select digits or characters from virtual keyboards shown in the HMD. For the alhpanumeric password a virtual keyboard, as  used by Boletsis et al.~\cite{Boletsis:2019} was presented to the users within the VR (shown in Figure~\ref{fig:keyboard}). Head-gaze based selection using a clicker has been suggested as the most preferred method for head-based gaze selection\cite{Pathmanathan:2020}. Another possibility is the use of built-in eye-trackers for selection~\cite{Blattgerste:2018}. However, most commercial VR HMD's do not provide built-in eye-tracking capabilities. To allow users to submit their response while wearing the HMD, we use SmartVR Pointer Gaze-based selection (SVRP2), a method that allows for selection using head-based gaze selection in VR using the smartphone as a clicker to confirm selections~\cite{SVRP:24}. Similar to Condition 1, this condition enables users to perform the whole authentication process relying solely on the VR HMD and their smartphone. A use case for this type of interaction is when a VR user is watching videos, browsing TV channels, or selecting a movie to watch, using their gaze for pointing and selection of their entertainment choice and then needs to confirm their identity to override a parental control. The system then generates a checkers-style challenge. Then, the user can use the smartphone to solve the challenge shown in VR using gaze-based selection, after which the user would have completed the  whole process using only their HMD and their smartphone.  

\subsubsection{Condition 3: Phone1\_VRC2 (a.k.a. the baseline)}
In this case the smartphone is used as the first device for the authentication, i.e., the device that shows the user the challenge to be solved (step 1, Figure ~\ref{fig:Phone1challenges}, right panel), and the HMD is used as the second device for the authentication, i.e., the device where the user enters and submits their solution to the challenge (step 2, Figures~\ref{fig:challenges_condition3}~and~\ref{fig:keyboard}). This is  similar to condition 2 above, with one major difference: the standard VR controller provided by the manufacturer of the HMD (VRC2) is used as the input mechanism within the VR. In this condition the user holds the smartphone with one hand (to see or revise the challenge) and the VR Controller with the other hand (as shown in Figure \ref{fig:challenges_condition3}A). This way, the user can  check the smartphone while solving the challenge using the VR controller. This is considered our baseline condition because most commercial VR HMDs come with VR controllers that are used for the purpose of selection, navigation, and other interactions in VR and most people make use of two-factor authentication by receiving text messages on their smartphones and entering the code on the application that originated the authentication request.
In a commercial application, the CAPTCHA, Numeric, and Password challenges could be submitted via text messages without having any app installed on the smartphone, except for the checkers challenge. We chose to use a smartphone app across all challenges for experimental consistency between conditions, making sure all tasks provided a consistent look and feel. In a user scenario, a gamer could be using the VR controller while playing as usual, and may receive a text message asking to confirm their identity to purchase an asset that is particularly expensive or to override a budgetary control that might have been set on the gamer's account preferences, in which case they can continue to use the VR controller to respond to the challenge presented in VR.

\subsection{Overview}
The traditional two-factor authentication most people are familiar with today is modeled by the conditions Phone1\_VRC2 (our baseline condition) and Phone1\_SVRP2, under which users would be using an HMD device hosting the VR app in lieu of the PC or device that would usually trigger the authentication process if there was no VR, then receive the authentication code in the supporting device (the smartphone), look at the validation challenge in the smartphone, and address it in the VR while wearing the HMD. In the reverse condition HMD1\_Phone2, the client app on the phone merely captures the user’s response to the challenge posed within the VR and submits the response to the server application after the user sees the challenge to be complete while inside of the VR. In all cases, the application that triggers the start of the authentication protocol is the main (server) application running in the VR system. 

Given the challenge types and conditions described above, we have designed a 4x3 experiment with four tasks or challenges (namely the CAPTCHA, Checkers, Numeric and Password challenges) and three conditions for NRXR authentication delivery, for a total of twelve authentication options. We then performed a user study to evaluate the feasibility and usability of each option and to measure users performance on each challenge per condition. 



\subsection{System Hardware and Software}
The system was implemented using the Unity 3D Game Engine, a widely recognized platform for the development of interactive applications and virtual environments. Furthermore, Unity's compatibility with smartphone development frameworks, such as Android SDK, allows for seamless incorporation of smartphone capabilities, including touch inputs and mobile-specific functionalities, such as tapping, drawing, or processing input from virtual keyboards.

\begin{figure*}[htbp]
    \centering
        \begin{minipage}{0.37\textwidth}
        \centering
        \includegraphics[width=\linewidth]{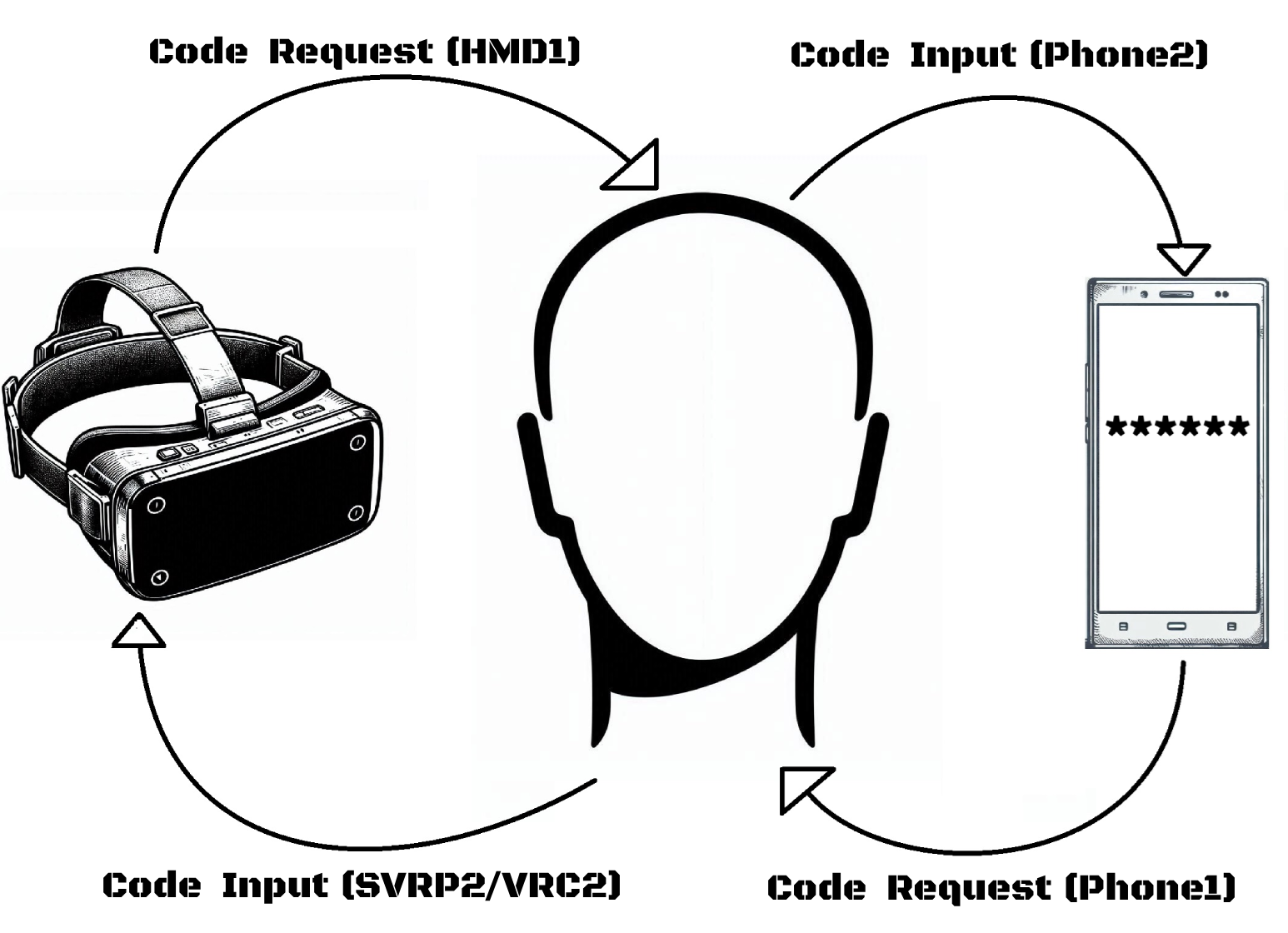} %
        \subcaption{Interaction modalities}
        \label{sketch}
    \end{minipage}
        \begin{minipage}{0.29\textwidth}
        \centering
        \includegraphics[width=\linewidth]{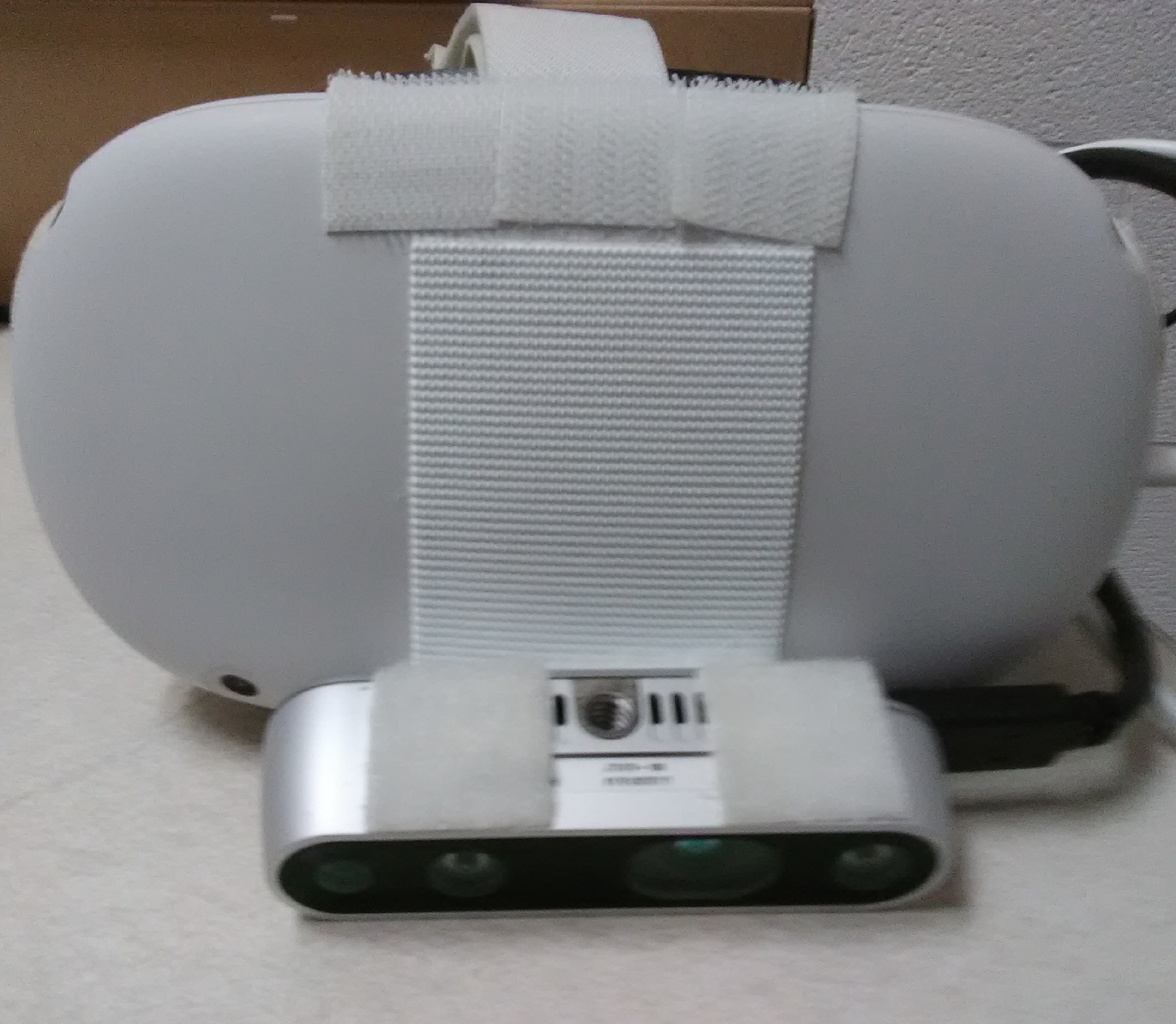}
        \subcaption{Front view}
        \label{front}
    \end{minipage} \hfill
        \begin{minipage}{0.30\textwidth}
        \centering
        \includegraphics[width=\linewidth]{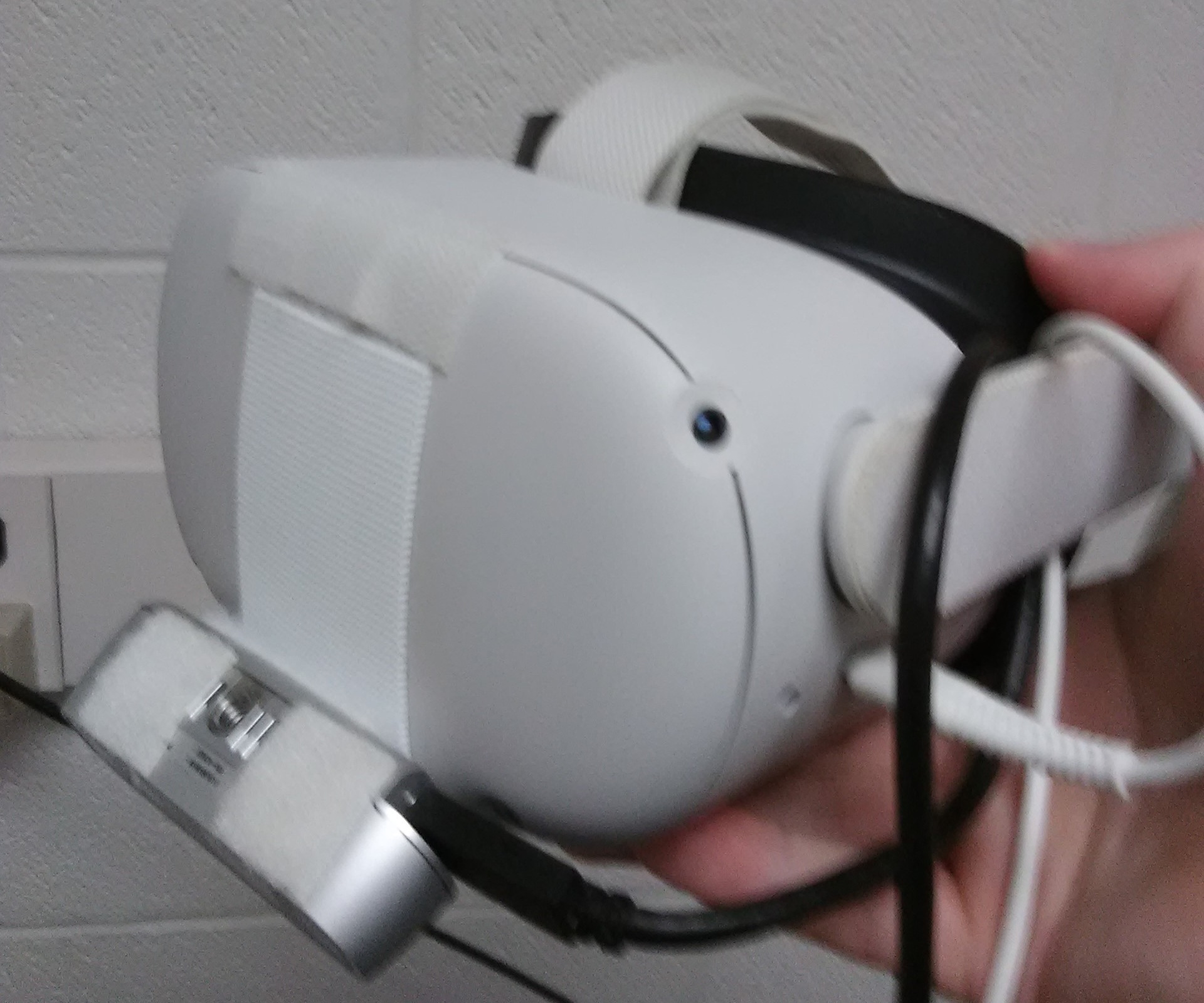} %
        \subcaption{Side view}
        \label{side}
    \end{minipage}
    \label{cameraAttached}
    \caption{Experimental Setup: a) Two interaction modalities - the code is requested by the HMD and the input is provided with the Smartphone, and vice versa; b) the front view displays the camera mounted on the headset, with the camera's position relative to the headset being adjustable.; c) Side view demonstrating the dynamic adjustment of the camera angle, oriented to facilitate a comfortable position for the user to hold the phone in front of the camera.}
    
\end{figure*}

The hardware and software utilized in this implementation included:
    
\begin{itemize}   
     \item \textbf {Meta Quest 2}
     Featuring a resolution of 1832 x 1920 pixels per eye and a refresh rate of 90 Hz, this headset provides a high-quality visual experience essential for immersion in virtual environments.
    \begin{itemize}
     \item Software Package Requirements: The integration of the Oculus PC app for Meta Quest Link establishes a seamless connection between the headset and the Unity engine. This software enables efficient data transfer and real-time rendering, minimizing latency and maximizing responsiveness in interactions.
     \end{itemize}

    \item \textbf{Intel RealSense and  Technology Developer Kit (SR300)}. This depth-sensing camera was incorporated to enhance close-range depth perception, a critical feature for user interactions within a VR context. The camera’s specifications—color resolution of 1920x1080 at 30 frames per second, along with an operating range of 0.3m to 2m—enable precise spatial awareness and tracking of user movements. Such capabilities are particularly important in tasks requiring accurate depth recognition, as they allow the system to interpret user actions in real time and respond appropriately, thereby facilitating a more intuitive interaction model~\cite{Taejin}.
    Depth Field of View: H: 73, V: 59, D: 90. Auto-exposure: Off. Brightness level: 350 (setup inside Unity).

    \item 
    \textbf{ZTE Z557BL Smartphone}: this device has a touchscreen resolution of 480 x 854 pixels, along with basic processing capabilities suitable for the application’s requirements. The Smartphone has 1.0 GB RAM and 8 GB storage. Android version: 8.1.0. Dimensions: 14.53 x 7.19 x 0.91~cm.
    \item \textbf{Software  requirements:}:We use SteamVR Runtime for Windows, along with Unity-compatible Intel RealSense SDK 2.0, to ensure cohesive operation among all hardware components within the Unity environment. We selected version of Unity (2020.3.25f1).
\end{itemize}

Figure \ref{sketch} illustrates the two-way interaction involving the HMD, the smartphone, and a human operator. In different scenarios, the request for the code may be initiated by either the HMD or the smartphone, while the second device is used for the code input. Figures \ref{front} and \ref{side} show front and side view of an experimental setup with the depth-sensing camera attached to the headset, and can be easily tilted to adjust the camera view angle for more convenient usage, depending on user's height, arm length, and preferred phone positioning.

\subsubsection{Blending of the RGBD camera feed}

In this implementation of NRXR, the IntelRealSense RGB-D camera is mounted on the Meta Quest 2 headset. The Intel RealSense camera combines a depth sensing camera and a regular video camera. In Unity, the RGBD stream from the camera is projected as a RGBD texture on a transparent window placed in front of the viewer. To blend the objects in the near range with the VR imagery, we use a fragment shader that takes in the input from the camera as an RGBD stream and adjusts the transparency for each pixel before sending the image to the display. If the depth value for a given pixel goes beyond a certain distance threshold ($\sim$120cm), we set that pixel to be completely transparent/pass-through, so that it does not block the view of the VR imagery behind the transparent window, allowing users to see most of the VR environment. A smoothing function for the alpha value is added for pixels near the distance threshold to ensure objects fade in and out smoothly while all pixels that are below the threshold become opaque. 

\subsubsection{Preventing overexposure during smartphone display  capture}

\begin{figure}[h]
    \centering
  \includegraphics[width=\columnwidth]{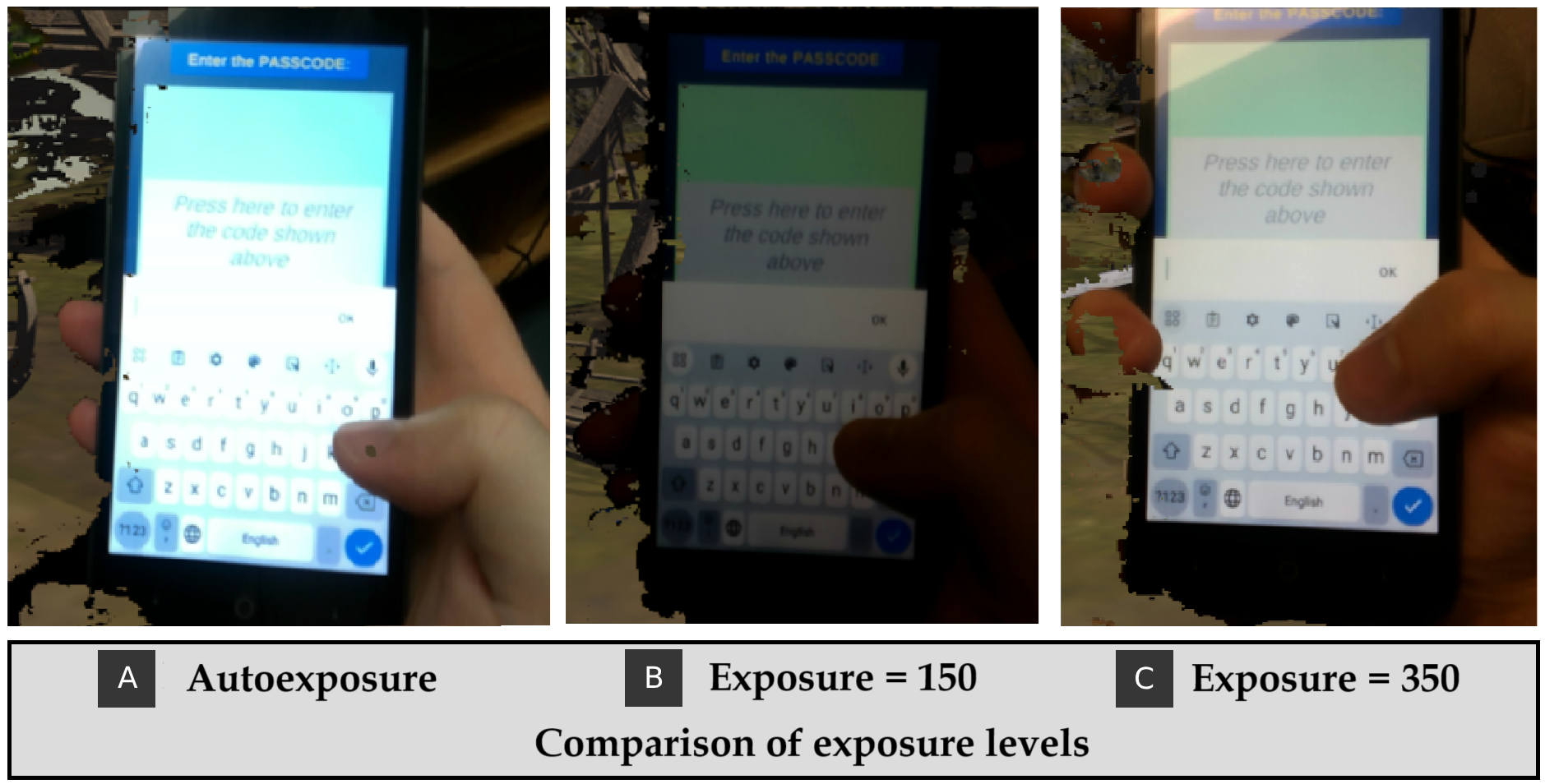}
  \caption{Adjusting exposure. A) Illustrates that pass-through video with autoexposure turned on often results in overexposure  of the smartphone screen's image; A) and B) show that too high or too low exposure levels cause the image to darken or brighten so much that it becomes nearly impossible to read its contents; C) Appropriate adjustment of the exposure level allows users to appreciate the screen contents.}
  \label{fig:exposure}
\end{figure}

To see the smartphone from within the HMD, the user has to place the smartphone in front of the HMD so that the phone is in front of the RGBD camera. Most cameras come with a dynamic auto-exposure function which automatically adjusts exposure according to the brightness levels of the captured scene to ensure video frame images are not too dark or bright under varying lighting conditions. This auto-exposure feature usually performs well for most applications, however when bright screens (such as the screen of a smartphone) are present in the captured video, there is often overexposure in the regions where the bright object is shown on the captured image. In the case of smartphones, the unintended consequence is that most of the portion of the video frames that capture the smartphone screen may become too bright and virtually blank due to overexposure. This tends to happen even more often when the room illumination is low and when users bring the smartphone screen closer to the HMD. Under these circumstances, users would have a hard time reading instructions and using the smartphone inside the VR. To deal with this issue we turned off auto-exposure on the RGB-D camera and set it to a low level of exposure so that the text and other content of the smartphone screen can be appreciated by the users. As illustrated in Figure \ref{fig:exposure}, adjustments of around 350±100 units  of exposure for the Intel Real Sense camera within the Unity profiler provides the best results for visibility, ensuring that the smartphone screen remains readable. An unintended consequence is that a small portion  of the surroundings of the smartphone become darker. In the case of the proposed setup, this is usually not problematic, since the focus is not on capturing surrounding objects but to allow the user to focus on the information presented on the smartphone screen. 

\section{Experimental design}
After ethics approval was received from an Academic Committee on Ethics in Human Research, we performed a 4x3 within-subject user study with N=30 participants in a VR scenario emulating a walk trough an old town dotted with teleporting markers. 
A pre-test questionnaire was conducted before the experiment to determine the demographic characteristics of the experiment participants. 

After completing the demographic questionnaire and signing the consent form for the data collection, the participants received a short set of instructions and proceeded to start the experiment.
Figures \ref{fig:challenges_condition1} to \ref{fig:challenges_condition3} illustrate the VR scenarios and conditions that we employed in our study. Participants were recruited through email distribution lists and postings in social media.  Before commencing the study,  participants were required to fill out an ethics consent form. Participants then went through 15 rounds through the VR environment using a VR controller to navigate using teleporting markers to move through  the VR environment. After that, they completed another 15 rounds using a smartphone and the SmartVR pointer technique~\cite{SVRP:24} to go through the environment in the same way as with the first time. Then, participants were randomly assigned to one group in which they would be exposed to all three conditions in different order (six different orders were possible, so each group had five participants). For each condition, participants started with one trial round (not measured). After the trial round, each participant performed 5 measured rounds, plus one round for collecting feedback, completing 7 rounds per condition in total. During the completion of each round, users found the challenges in the following order: CAPTCHA, Numeric, Checkers, and Password. Users moved from one challenge to the next using teleporting locations which directed them to the next challenge. The teleporting task provided a washout period between challenges. We collected the time it took participants to complete each challenge, from the moment participants arrived to the challenge until they solved it, excluding the time that participants spent teleporting  from one challenge to the next. For instance, if a participant failed to travel to the start of a challenge the first time and then succeeded the second time, thereby taking longer than others, it would not affect the task completion time recorded for the challenge at where the participant arrived. 
We also collected the number of unsuccessful attempts for each challenge. 
On the seventh round, we collected participants feedback using a 7-point Likert scale to answer a user experience questionnaire containing the following questions after each challenge:

\begin{itemize}
    \item How much did you like this way of authenticating? \\
1  (I did not like it at all) 2  3  4 (Neutral) 5 6 7 (I liked it very much)
    \item On a scale from 1 to 7, How effective did you find this way of authenticating? \\
1 (Not effective at all) 2  3  4 (Neutral) 5 6 7 (Very effective)
    \item On a scale from 1 to 7, How easy to use did you find this way of authenticating? \\
 1 (Not easy at all) 2  3  4 (Neutral) 5 6 7 (Very easy)
\end{itemize}

At the end of the experiment, we additionally collected unstructured feedback related to the  participants' experience using a paper-based questionnaire with open-ended questions. The questions included the request to provide users' feedback on whether they preferred any particular methods, challenges and input type conditions provided to them during the experiment; what symptoms the VR experience may have caused, and what tasks or interactions they found to be easiest and hardest.

\subsection{Data Analysis}

To test whether all six order groups' composition in terms of participants' gender, age, occupation and previous HMD experience was as expected from a random participant group allocation, we performed Pearson’s Chi-square tests with \emph{{p}}-values calculated by Monte Carlo simulation with 100,000 replicates. To evaluate differences on participants' task completion time and number of clicks between  conditions, we performed ANOVA analyses for each of the four tasks. In the ANOVA analyses, the effects of the order and round number were considered. We used Tukey’s test as a post-hoc analysis for pair-wise comparison of the means between conditions, order and rounds. To test whether participants’ responses to the Likert-scale questions in the user experience questionnaire differed among conditions, we performed pair-wise Wilcoxon tests with false discovery rate (FDR) correction for multiple testing.  All statistical data analyses were conducted in R (version 4.3.1) and plots were created using R data visualization package ggplot2 (version 3.4.3).


\section{Results}

\subsection{Participant Demographics}


Out of the 30 participants, 40\% of them (12/30) were aged 18 to 24, 56.67\% of participants (17/30) were aged 24 to 48, and one participant was aged 48 to 65 (3.33\%). About 36.67\% (11/30) were computer science students, both graduate and undergraduate, while another 40\% of were non-computer science students (12/30); the 7 others (23.33\%) were either staff, faculty or alumni. One third of participants (10/30) identified as female and two thirds identified as male (20/30).

Almost half of the participants (14 people, 46.67\%) had never used and HMD. A third (33.33\%, 10/30) had used an HMD for less than 1 month in total, while 10\% (3/30) had between 1 and 6 months experience with HMDs; another 6.67\% (2/30) had significantly more experience using HMDs (6 months to 2 years); one participant preferred not to answer the question regarding HMD usage experience.

\subsection{Performance Metrics}

\begin{table}
    \centering
    \small 
    \setlength{\tabcolsep}{4pt}
    \begin{tabular}{c|l|l|l|l}
    \toprule
        \textbf{Condition} & \textbf{CAPTCHA} & \textbf{Numeric} & \textbf{Checkers} & \textbf{Password} \\
        \midrule
        HMD1\_Phone2 & $17.76 \pm 10.46$           & $\underline{\mathbf{10.68 \pm 3.27}}$  & $13.63 \pm 5.18$          &  $33.45 \pm 19.22$\\
        Phone1\_SVRP2 & $12.92 \pm 4.16$           & \underline{$12.47 \pm 4.79$}           & $14.47 \pm 6.50$          &  $\mathbf{28.35 \pm 9.33}$ \\
        Phone1\_VRC2  & \underline{$\mathbf{12.41 \pm 5.02}$}   & $13.90 \pm 5.92$           & $\mathbf{13.44 \pm 9.33}$ &  $36.32 \pm 18.05$ \\
        \midrule
        Average $\pm$ sd & $14.36 \pm 7.5$  & $\mathbf{12.35 \pm 4.95}$ &   $ 13.85 \pm 5.29$ &  $32.71  \pm 16.44$  \\
          \bottomrule
    \end{tabular}
    \caption{Mean completion times (in seconds) and standard deviation per condition for each of the four challenges. The lowest mean completion time per challenge and the lowest overall average completion time are highlighted in Bold. The challenge with the lowest mean completion time per condition is underlined.}
    \vspace{-20pt}
    \label{tab:TimeTable}
\end{table}

\begin{table}[htb]
    \centering
    \small 
    \setlength{\tabcolsep}{2pt}
    \resizebox{1.05\columnwidth}{!}{
    \begin{tabular}{c|r|r|r|r|r|r|r|r}
    \toprule
    \textbf{Factor} & \multicolumn{2}{c}{\textbf{CAPTCHA}} & \multicolumn{2}{c}{\textbf{Numeric}} & \multicolumn{2}{c}{\textbf{Checkers}} & \multicolumn{2}{c}{\textbf{Password}} \\
        & F-value & p-value  & F-value & p-value  & F-value & p-value  & F-value & p-value \\
    \midrule
    Condition  & 30.02 & $\mathbf{5.99\times10^{-13}}$ & 17.45 & $\mathbf{5.09\times10^{-8}}$  & 1.69 & $ 0.185$ & 9.69 & $\mathbf{7.62\times10^{-5}}$\\
    Round      & 12.13 & $\mathbf{2.31\times10^{-09}}$ & 3.17 &  $0.014$              & 3.49 & $\mathbf{ 0.008}$ & 1.90 & 0.11\\
    Order      & 15.28 & $\mathbf{3.83\times10^{-07}}$ & 1.75 & $ 0.175$     & 7.14 & $\mathbf{0.0009}$ & 6.88 & \textbf{0.001}\\
     \bottomrule
    \end{tabular}
    }
    \caption{ANOVA results of completion times per challenge. P-values less than 0.01 are highlighted.}
    \vspace{-20pt}
    \label{tab:ANOVA_results}
\end{table}

Overall, participants took the longest to complete the Password challenge, and were the quickest to solve the Numeric challenge (Table~\ref{tab:TimeTable}). The independent variable that most significantly affected the overall completion times was the challenge (ANOVA p-value $< 2\times10^{-16}$) followed by the interaction between challenge and condition (ANOVA p-value $= 2.72\times10^{-15}$). The effect of the condition on the overall completion times was significant with an ANOVA p-value of $= 0.0002$.  Based on the ANOVA results per challenge, we found that in all challenges except the Checkers challenge, the factor that most affected the completion time of the participants was the condition (Table~\ref{tab:ANOVA_results}). The effect of the order in which the conditions were presented to the participants or the round number varied substantially between the challenges. For instance, these two factors (round and order) were found statistically significant for the CAPTCHA and Checkers challenges; while having a smaller effect on the Numeric and Password challenge. This indicates that one needs to take the authentication challenge/task into consideration when selecting a configuration modality.

\begin{figure}[h]
    \centering
   \includegraphics[width=\columnwidth]{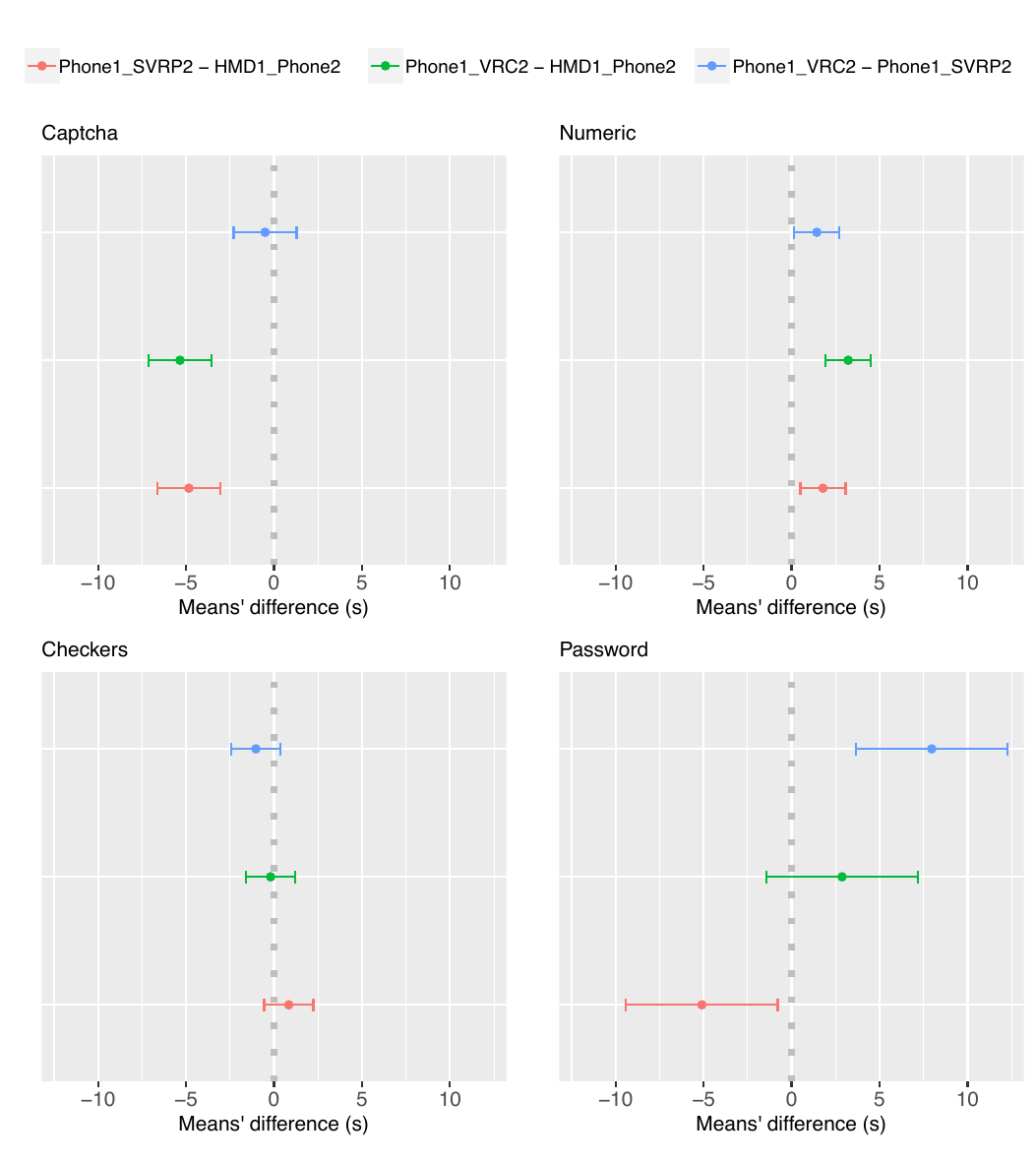}
    \caption{95\% confidence intervals of the pairwise differences in mean completion time between conditions for all four challenges. Circles indicate the mean difference. Dashed vertical gray line indicates the point of no difference between the means. The farthest the confidence interval is from the dashed vertical line, the more statistically significant is that difference. Differences are in seconds.}
    \label{fig:MeanTimeCompletionDifferences}
\end{figure}

As expected, and based on the ANOVA results, there were statistically significant differences in mean completion time per condition for the CAPTCHA, Numeric and Password challenges (Figure~\ref{fig:MeanTimeCompletionDifferences} and Table~\ref{tab:TimeTable}). The largest observed differences in participants' performance were the following: in the Password challenge, participants were on average 8 seconds faster in the Phone1\_SVRP2 condition than in the Phone1\_VRC2 condition. In the Numeric challenge, participants were on average 3.2 seconds faster in the HMD1\_Phone2 condition than in the Phone1\_VRC2  condition. Finally, in the CAPTCHA challenge, participants were 5.3 and 4.8 seconds slower in the HMD1\_Phone2 condition than in the Phone1\_VRC2 and Phone1\_SVRP2 condition, respectively.

\begin{figure}[h]
    \centering
    \includegraphics[width=\columnwidth]{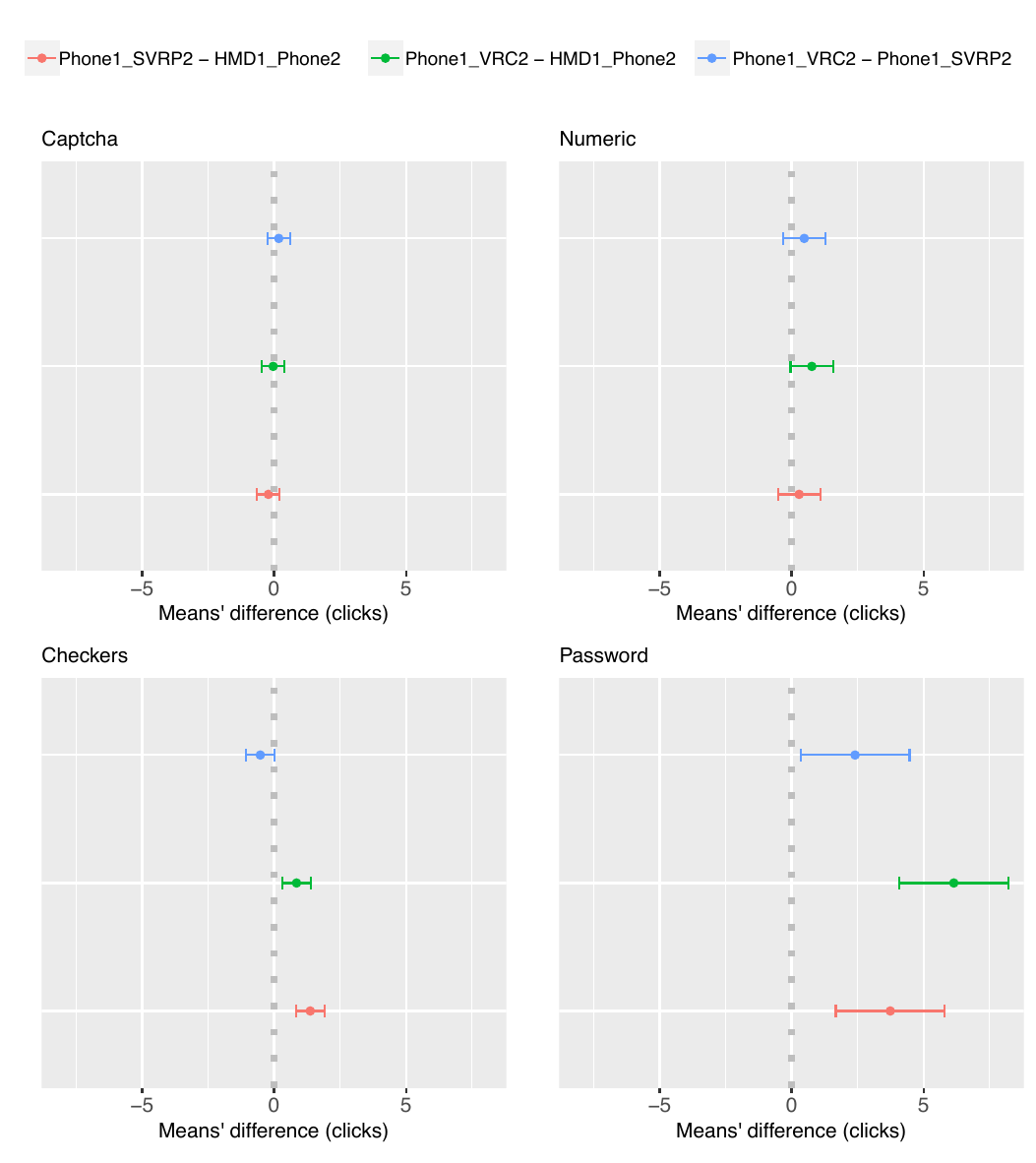}
    \caption{95\% confidence intervals of the pairwise differences in mean number of clicks between conditions for all four challenges. Circles indicate the mean difference. Dashed vertical gray line indicates the point of no difference between the means. The farthest the confidence interval is from the dashed vertical line, the more statistically significant is that difference. Differences are in number of clicks.}
    \label{fig:MeanClicksDifferences}
\end{figure}

As mentioned previously, the minimum number of clicks required to complete any of the challenges was six. On average, participants required  $7.96 \pm 2.97$, $8.3 \pm 2.04$, $8.8\pm 1.56$, and $13.97 \pm 7.99$ clicks to complete the Numeric, Checkers, CAPTCHA and Password challenge, respectively. The number of clicks was not significantly affected by the condition, order or round in the CAPTCHA and Numeric challenges. Condition was  the only significant factor in the Checkers (F value 18.72, p-value $1.57\times10^{-8}$) and in the Password (F value 24.93, p-value $5.55\times10^{-11}$) challenges. In the Checkers challenge, participants needed on average 1.4 and 0.85 more clicks to complete the challenge in the Phone1\_SVRP2 and Phone1\_VRC2 conditions, respectively, than in the HMD1\_Phone2 condition (Figure~\ref{fig:MeanClicksDifferences}). In the Password challenge, participants needed on average 3.74 and 6.15 more clicks to complete the challenge in the Phone1\_SVRP2 and Phone1\_VRC2 conditions, respectively, than in the HMD1\_Phone2 condition (Figure~\ref{fig:MeanClicksDifferences}). 

\begin{table}[htb]
    \centering
    \small 
    \setlength{\tabcolsep}{2pt}
    \resizebox{1.05\columnwidth}{!}{
    \begin{tabular}{c|c|c|c|c||c}
    \toprule
        \textbf{Condition} & \textbf{CAPTCHA} & \textbf{Numeric} & \textbf{Checkers} & \textbf{Password} &   \textbf{Average} \\
        \midrule
HMD1\_Phone2      &   85\%      &                     97\%          &    \textbf{92}\%                 &              88\% & 90\%\\
Phone1\_SVRP2     &  94\%        &                  \textbf{ 99\%  }          &             89\%               &      \textbf{ 91\%} & \textbf{93\%}\\
Phone1\_VRC2      & \textbf{96\%}      &                    93\%          &              91\%              &               87\% & 91\% \\
\midrule
Average $\pm$ sd  &  $91.67\%\pm 5.86\%$   &       $\mathbf{96.33\%  \pm 3.06\%}$   &       $90.67\% \pm 1.53\%$       &     $88.67\% \pm 2.08\%$ & 91.83\%\\
        \bottomrule
    \end{tabular}
    }
    \caption{Success rate per condition for each of the four challenges. The highest success rate per challenge and the highest overall average success rate are highlighted.}  
    \vspace{-20pt}
    \label{tab:successRate}
\end{table}

We tracked the number of unsuccessful attempts made by the participants and calculated the success rate for each challenge and condition combination as the percentage of successful attempts over total attempts  (Table~\ref{tab:successRate}). Overall success rates averaged from 90\% to 93\% for each condition. The challenge with the highest success rate (96.33\%) was the Numeric challenge, followed by CAPTCHA and Checkers. The condition with the highest  overall success rate  was Phone1\_SVRP2 (93\%). The challenge that had the lowest success rate was the Password challenge (88.67\%).

Unsurprisingly, there was a significant positive correlation between completion time and number of clicks. This correlation was the strongest for the Password challenge (Spearman's $\rho$ 0.53, p-value < $2.2\times10^{-16}$) and the slightest for the CAPTCHA challenge (Spearman's $\rho$ 0.15, p-value 0.002).

\begin{figure}[h]
   \centering
        \includegraphics[width=\columnwidth]{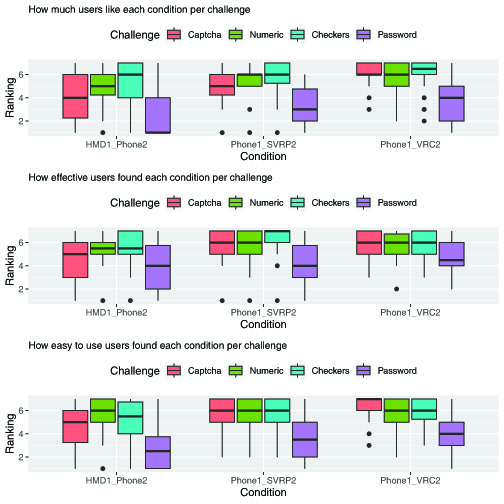}
    \caption{Distribution of Likert-scale scores given by participants to each condition per challenge based on how much they liked the condition, how effective they found it and how easy to use they perceived it. In the ranking, 7 is best, 4 is neutral and 1 is worst. A horizontal line inside the box indicates the median score, and the box height indicates the interquartile range (IQR).}
    \label{fig:Feedback}
\end{figure}

\subsection{Participant Feedback}

\subsubsection{Structured Feedback Analysis}

In terms of participants' structured feedback (using Likert scales), participants liked more the Phone1\_SVRP2 and Phone1\_VRC2 conditions (FDR-adjusted Wilcoxon rank's p-value 0.007 and $1.9\times10^{-5}$, respectively) than the HMD1\_Phone2 condition (see Figure~\ref{fig:Feedback}). Similarly, participants found more effective the Phone1\_SVRP2 and Phone1\_VRC2 conditions (FDR-adjusted Wilcoxon rank's p-value 0.003 for both) than the HMD1\_Phone2 condition. With the same trend, participants perceive the Phone1\_SVRP2 and Phone1\_VRC2 conditions (FDR-adjusted Wilcoxon rank's p-value 0.003 and 0.0001, respectively) easier to use than the HMD1\_Phone2 condition. Participants' scores for the Phone1\_SVRP2 and Phone1\_VRC2 conditions were comparable.

Regarding the challenges, the participants liked less the password challenge (FDR-adjusted paired Wilcoxon rank p-values $\le 2.4\times 10^{-13}$) than any of the other three challenges. Participants liked more the checkers challenge than the CAPTCHA and numeric challenges (FDR-adjusted paired Wilcoxon rank p-values of 0.009 and 0.023, respectively), and they liked slightly more the numeric challenge than the CAPTCHA challenge (FDR-adjusted paired Wilcoxon rank p-values of 0.067). Similarly, participants found less effective the password challenge (FDR-adjusted paired Wilcoxon rank p-values $\le 1.4\times 10^{-7}$) than any of the other three challenges, and the checkers challenge more effective than the CAPTCHA and numeric challenges (FDR-adjusted paired Wilcoxon rank p-values of 0.028 and 0.040, respectively). There was no statistically significant difference in the perception of effectivity by the participants between the numeric and CAPTCHA challenges. Finally, the participants found the password challenge more difficult (FDR-adjusted paired Wilcoxon rank p-values $\le 6.7\times 10^{-14}$) than any of the other three challenges. There was no statistically significant difference in how difficult the participants found the other three challenges.

\subsubsection{Unstructured Feedback Analysis}

In terms of unstructured feedback, not all participants expressed a particular preference for a certain combination of a challenge and condition, however, the most frequently mentioned case as the best combination was resolving the checkers challenge during condition 2, Phone1\_SVRP2. Exemplifying this view, a participant commented: ``reading the codes from the smartphone's screen seemed the most efficient to me due to the possibility to simultaneously look and read the codes and enter the answer with head movement. Liked the idea of authentication via ``checkers'' system, it was the most entertaining one''. 



\begin{table}[htb]
    \centering
    \small 
    \setlength{\tabcolsep}{2pt}
    \resizebox{1.05\columnwidth}{!}{
    \begin{tabular}{c|l|l|l|l|l}
    \toprule
        \textbf{Category} & \textbf{CAPTCHA} & \textbf{Numeric} & \textbf{Checkers} & \textbf{Password}& \textbf{Total} \\
        \midrule
        Best challenge  & 2.38\%& 4.76\%& 14.29\%& 0.00\%& \textbf{21.43}\%\\
        Good challenge  & 7.14\%& 11.90\%& 11.90\%& 2.38\%& \textbf{33.33}\%\\
        \midrule
        Bad challenge   & 7.14\%& 7.14\%& 0.00\%& 19.05\%& \textbf{33.33}\%\\
        Worst challenge & 2.38\%& 0.00\%& 0.00\%& 9.52\%& \textbf{11.90}\%\\
        \midrule
        \textbf{Balance} & \textbf{0.00\%}	&\textbf{9.52\%}	&\textbf{26.19\%}	& \textbf{-26.19\%} &   \textbf{9.52\%}  \\
          \bottomrule
    \end{tabular}
    }
    \caption{Summary of participants' comments  regarding the challenges encountered during the experiment. The last row shows the balance of the percentages of  positive comments minus the negative ones.}
    \vspace{-20pt}
    \label{tab:tasks_pref}
\end{table}
Participants expressed their preferences regarding the best and worst challenges provided during the experiment by answering the unstructured feedback section in the post-questionnaire.  The results of the unstructured feedback regarding the challenges is summarized in Table~\ref{tab:tasks_pref}. As shown in the table, comments were read and sorted in 4 categories per challenge. We then calculated the percentages of comments in each cell and divided them over the total of comments (42 comments, 23 positive and 19 negative in total). 
By far, the most positively commented challenge  was the Checkers challenge, with 26\% of all comments mentioning it as either a good challenge or the best challenge and no negative comments. The runner-up was the numeric code with 16.67\% of comments being positive about it  and 7.14\% being negative. 
The CAPTCHA challenge got mixed replies evenly, with 9.52\% of  positive comments and 9.52\%  negative comments. 
The Password challenge caused users to struggle the most, as evidenced by 28.57\% of comments being negative about it and only 2.38\% of comments being positive. To exemplify, one of the participants commented: ``With respect to entering the code from the keypad of the phone, I almost got the feeling that my grandpa gets while he enters a text message''. 

\begin{table}
    \centering
    \small 
    \setlength{\tabcolsep}{2pt}
    \resizebox{1.05\columnwidth}{!}{
    \begin{tabular}{c|l|l|l|l}
    \toprule
        \textbf{Category} & \textbf{HMD1\_Phone2} & \textbf{Phone1\_SVRP2} & \textbf{Phone1\_VRC2} & \textbf{Total} \\
        \midrule

Mostly Positive & 0.00\% & 21.43\% & 9.52\% & \textbf{30.95\%} \\
Positive & 7.14\% & 14.29\% & 14.29\% & \textbf{35.71\%} \\
\midrule
Negative & 9.52\% & 2.38\% & 7.14\% & \textbf{19.05\%} \\
Mostly Negative & 9.52\% & 0.00\% & 4.76\% & \textbf{14.29\%} \\
\hline
\textbf{Balance} & \textbf{-11.90\%} & \textbf{33.34\%} & \textbf{11.91\%} & \textbf{33.32\%} \\

          \bottomrule
    \end{tabular}
    }
    \caption{Summary of participants' comments  regarding the conditions encountered during the experiment. The last row shows the balance of the percentages of  positive comments minus the negative ones.}
    \vspace{-25pt}
    \label{tab:conditions_pref}
\end{table}

We performed a similar analysis for the three conditions. Regarding these, we registered 42 comments in total, with twice as many positive than negative entries (28 positive vs. 14 negative).  Participants overwhelmingly preferred the  Phone1\_SVRP2 condition over the two other conditions, with 35.71\% of all comments liking most or some of its features and only 2.38\% disliking some aspect of it.
The second most preferred condition was Phone1\_VRC2 with 23.81\% comments liking some or most of it, but 11.9\% disliking some or most of it. The HMD1\_Phone2 condition was the least preferred with 7.14\% of comments liking some or most of it, but 19.05\% disliking some or most of it, with an overall negative balance between positive and negative comments, which may stem from the difficulties participants found when completing the CAPTCHA and Password challenges. See Table~\ref{tab:conditions_pref} for the detailed results.

\subsection{Summary of Results}
When considering the four challenge types, users were on average fastest with the numeric PIN challenge, followed closely by the checkers challenge and the CAPTCHA challenge, with the keyboard password a distant fourth place. Across the structured feedback, users preferred the checkers challenge the most, followed by the numeric PIN challenge. In terms of unstructured feedback, The CAPTCHA received equal amount of positive and negative remarks, and the alphanumeric password received most negative comments.

The effect of the condition in participants' completion time varied across challenges. While the password challenge seems to be the least suitable to be used in a VR environment overall, participants completed this challenge much faster in the Phone1\_SVRP2 condition, highlighting the convenience of gaze-based interaction supported by a smartphone. For  the checkers-style challenge, the condition used did not impact users' performance, which speaks in favour of it, as it offers consistent performance across conditions. For the numeric PIN challenge, participants completed the challenge  the fastest in the HMD1\_Phone2 condition and was laso the fastest for the Phone1\_SVRP2 condition. For the CAPTCHA challenge, the Phone1\_SVRP2 and Phone1\_VRC2 conditions allowed participants to complete the challenge clearly faster than in the HMD1\_Phone2 condition, which may be explained by the fact that it is easier to recognize the tile's contents when they are shown in the VR than when they are shown via NRXR. In general, users performed fastest under Phone1\_SVRP2, and perceived more positively the Phone1\_SVRP2 and Phone1\_VRC2 conditions than the HMD1\_Phone2 condition.

\section{Discussion}
 With respect to Research Question 1 (RQ1), we observed that all participants were able to complete all challenges under all three conditions. This is encouraging, given the variety in the authentication challenges presented. The success rates observed provide strong support for the view that it is in fact possible to use NRXR-ID to support users to authenticate their credentials via two-factor authentication, and that users are able to fully identify text, digits, images, and patterns without removing their HMDs. However, we detected significant differences between the different challenges and conditions overall, as described below.  
 
 With respect to RQ2, regarding which type of authentication challenges are most suitable to implement two-factor authentication in the VR context, our results indicate that the checkers challenge was among the most suitable for deployment, as it was the most liked, most effective and easiest to use among all the conditions. While some participants excelled in some challenges under certain configurations, checkers produced good results overall, was consistent across conditions in terms of average execution times, and was clearly the most preferred by users in the unstructured feedback results. The numeric PIN challenge was consistently a good alternative to the checkers challenge and has the advantage that is probably the method most users are familiar with, does not require a smartphone app in the Phone1* conditions, and allowed for the fastest average completion times among all four type of challenges as well as for condition Phone1\_SVRP2. In addition, the numeric code was the method with the highest average success rate and the highest success rate overall when combined with condition Phone1\_SVRP2. In the unstructured feedback, only the checkers challenge received more positive comments than the  numeric PIN code. These results might be explained in part by a novelty effect involving the checkers challenge. 
 
 The CAPTCHA challenge was third in performance and was ranked neutral to quite positive in user preferences, but it was least liked, found least effective, and least easy to use in condition 1 (HMD1\_Phone2). Even then, it had the third highest success rate when used in condition Phone1\_VRC2 and the fourth highest success rate when used in condition Phone1\_SVRP2, when considering all combinations of challenge and condition. We believe the reason for this is that it is harder to recognize the shapes in the  tiles using the NRXR technique than it is to recognize them when they are shown straight on the HMD. 
 
 From all four challenges it is quite clear that the password challenge is the least suitable for the second step of the authentication, i.e., the confirmation of the user's identity. Not only it involves the recognition of a larger set of letters and special characters, but each letter and special character occupies a smaller area in the display. This is specially true in condition HMD1\_Phone2. The password challenge is also the most complex to use, in that users had to switch between special, upper, and lower case key sets, so even though users were asked to enter six characters as in the other challenges, the complexity of the task was larger due to the need to switch between keyboard sets. In addition, the penalty for correcting a mistake was larger for keyboard users, since they had to delete some successfully entered characters until they returned to the character that might have been incorrectly entered in the first place. All these factors may explain why even when participants had a larger virtual keyboard in conditions 2 and 3 (Phone1\_SVRP2) and (Phone1\_VRC2), they still ranked password the lowest from among all four challenges. It can be argued that the password challenge is the most robust to a brute force attack, as it allows for approximately 100 billion configurations in this experimental setup, and even more with a larger number of characters. Since it is one of the most reliable forms of authentication in non-VR environments using 2FA~\cite{Jubur:2025}, we believe the password challenge could be used as the first line of authentication in VR, i.e., when the user first gains access to the XR environment, such that, after that, one of the three other methods could be used for confirmation of a user's identity when the need arises.
 
 With respect to RQ3, on whether it is most effective for users to present a challenge using the smartphone and have the challenge answered within the VR (conditions Phone1\_SVRP2 and Phone1\_VRC2), or the other way around (condition HMD1\_Phone2), results show that overall participants found it is less effective to present the challenges using the HMD first and have the challenges responded using the smartphone next (HMD1\_Phone2), and this is mostly due to the fact that both CAPTCHA and password challenges were found least effective under this condition. When the challenge was checkers or numeric, users found both  variants equally effective.  This is also why we suggest that the most suitable method for authentication is the checkers matching challenge and that the second most suitable is the numeric challenge, as they are more robust to the changes in the delivery condition. CAPTCHA was ranked generally higher in the Phone1 conditions (Phone1\_SVRP2 and Phone1\_VRC2) than in the HMD1\_Phone2 condition. Overall, the condition with the highest success rate was Phone1\_SVRP2 followed by Phone1\_VRC2. 
 
 With respect to RQ4, collecting  both structured and unstructured feedback was useful to obtain insights regarding the participants experiences, impressions and preferences. Overall, it is clear that the percentage of participants who preferred the checkers challenge was comparable to the percentage of participants who  disliked the  password challenge, so these techniques lie in opposite ends of the spectrum. The numeric challenge was the second most embraced challenge after checkers. The percentage of positive and negative views on the CAPTCHA challenge balanced each other across conditions.  With the exception of the password challenge, users found the three other challenges fairly easy to use irrespective from the interaction modality. In terms of the different conditions, a much larger proportion of participants had a positive impression of condition Phone1\_SVRP2 than about the  two other conditions.
 
 \subsection{Security Considerations and Potential Vulnerabilities}
 While 2FA offers increased security in non-VR settings, VR specific vulnerabilities such as potential gaze exploitation~\cite{Wang:2024} and de-anonymization attacks~\cite{Sabra:2024} may still remain in NRXR-ID, with the exception of an increased difficulty to insert a man-in-the-middle, or object-in-the-middle attacks~\cite{Fujita:2023,Lebeck} due to challenges for attackers in interfering with the video feeds of the HMD built-in cameras or with the user's smartphone. With respect to the vulnerabilities associated with the proposed use of NRXR, it is important that hackers and collaborators do not have access to the HMD's built-in camera feeds, to prevent them from seeing the user's smartphone through a network connection. This can be done by implementing a hardware and/or software lock, such as Visual Cryptography~\cite{Andrabi}, which decodes messages sent from machines for human observers to perceive. Ideally, this mechanism should be activated during authentication in such a way that it only allows the feed from the NRXR to be visible to the person wearing the HMD, while no one else except that person has a view of the NRXR. 
 
 With respect to the challenges explored in the study, it is worth noting that some participants reported that the CAPTCHA challenge was "relatively easy to guess". We believe that this impression may be caused by the fact that CAPTCHA tiles share information across multiple tiles, whereas the other challenges do not require any internal consistency  to solve the  challenge (e.g., knowing one digit of the six-digit PIN code does not help in knowing the other five, as all six digits are independent from each other). Some CAPTCHA users may identify a "theme" (i.e., a group of images that belong together) and could make a guess even without receiving the challenge. Having larger grids of tiles (e.g. 5x5) would help reduce  this issue. However, each tile then would become visually smaller (less screen real estate), and its contents may become harder to recognize, which would make the challenge harder to solve and potentially more frustrating for users. In addition, CAPTCHA may be potentially vulnerable to a brute-force attack combined with a machine learning attack, where machine learning could be used to identify themes in the tiles to make educated guesses as part of the brute-force attack. For these reasons, we believe the CAPTCHA challenge may be the most vulnerable to these type of attacks. With respect to the checkers challenge, in the experimental setup we used a 4x4 grid, providing  2$^{16}$ (or 65,536) different combinations. For increased robustness, the grid could be expanded to a 5x4 grid, where there would be  2$^{20}$ possible arrangements, which is a little more than a  million different combinations and would make the checkers challenge as hard to crack by brute force as the six digit numerical code, which provides 1 million options. Finally, the alphanumeric password was limited to 6 characters, whereas most current secure password guidelines recommend to create a password with at least 8 characters. Since the results clearly show that this was the least preferred option and the one that took the longest, one can only expect that increasing the length of the password to be provided to 8 characters will only make this option even less preferred by users and should take even longer. To address this issue, it has been suggested that such passwords be handled using password managers, which combined with 2FA would offer the most robust combination in non-VR settings~\cite{Tirfe:2021,Jubur:2025}. This might be also the case when using XR systems. However, in the XR device the password manager system would then need to be protected from misuse and the HMD would need to be revised for potential vulnerabilities as well~\cite{Sha:2025}.

\subsection{Limitations}
We have identified several limitations with the current setup. The first limitation is the presence of cables. The latest generation of HMDs almost completely liberate users from the need to remain connected to a computer or laptop.
Ideally, users should operate without a physical cable connecting them to a  workstation. We considered the possibility of using wireless cameras, but found that the delay between the user actions and the time these actions were reflected in the VR world would be a major limiting factor due to the transmission times. This issue may be revised once faster network protocols become the norm. The depth-sensing camera we used is not designed to be wireless. Our setup would require extra hardware additions to be able to be entirely wireless, adding to the current burden of wearing the HMD and the camera itself. Another limitation lies in the types of challenges that can be used for two-factor authentication in VR. There is a very large number of possibilities to implement two-factor authentication challenges in the virtual world. For example, the use of gaming-style 3D puzzles such as those involving more complex  challenges where users need to rotate an object in 3D space to discover a code to be used on the second device could be considered, as well as challenges involving Mixed-Reality juxtaposition of objects from the real and virtual world. The possibilities are vast, so we chose those which are well established, and leave others for future exploration.

\subsection{Future Work}
To achieve a more generally accessible implementation of NRXR-ID, we are exploring the possibility of using the built-in cameras in the HMD's to extract the depth field without the need of an external camera. Just until very recently, Meta has provided developers with access to the passthrough cameras of its high-end HMDs (Meta Quest 3 and Meta Quest Pro), opening the door to an implementation of NRXR in their systems. The Apple Vision Pro already allows users to see and operate their smartphones in well-illuminated environments. However, we were unable to use it for our studies, as it was  released just last year and it would take significant time to develop the study in the development environment of the Apple ecosystem. In addition, the Apple Vision Pro costs approximately 10 times more than a Meta Quest 3 and 5 times more than a Meta Quest Pro, which raises questions of general access to the device, how this would limit the availability of the solution to the enterprise segment of the XR market, and how this might reduce the impact or the availability of 2FA to the wider XR community.

In a separate line of enquiry,  we are working towards the implementation of NRXR without the use of a depth stream altogether.  Machine learning methods can be used to  perform real-time segmentation by detecting smartphones and users hands from RGB streams, which could be then be used to simulate or replace the need for depth estimation. These  modifications would provide the crucial benefit of  eliminating the need for the cable to connect the depth sensing camera. Another potential improvement would be the use of mid-air hand tracking (also known as hand-based interaction) to let users execute the second part of the two-factor authentication process without a VR controller. Even though this would also  eliminate the need for the VR controller when answering the authentication challenge in Phone1-style conditions, it would also remove the haptic feedback and the tangible nature of the setup, which is something that has been reported as having its own advantages in related literature~\cite{PhoneInVR2024,Zhang2021}.  Regarding gesture detection, it is worth noting that there is no single set of gestures for interacting in VR that has become established as an industry standard. For instance, the Apple Vision Pro relies solely on eye-tracking in combination with the pinch gesture, the Microsoft HoloLens relies on pinching, sliding and poking in the air for selection, and supports other gestures through their Mixed Reality Toolkit (MRTK), while the Meta Quest 3 supports its own set of gestures through their Interaction SDK, which supports both hand tracking and VR controllers. We are planning to design follow-up user studies comparing the most preferred methods discussed in this article with hand-tracking and hand-gestures based interaction techniques. As mentioned above, using the cameras already built in the latest generation of HMDs would yield  another potentially significant improvement. During the production of this article, Meta Quest HMD's do not provide developer access to the RGBD or the depth streams in a way that would allow us to implement NRXR without the external camera. Many HMD manufacturers cannot provide or do not allow software developers to have direct access to the video \& depth streams from the built-in cameras, citing potential privacy concerns.  The streaming data from the HMD could be a stereo video stream, from which the depth could be extracted or, if the HMD has built-in depth sensors, a depth video stream. Having access to the built-in hardware in the HMDs to implement NRXR would be useful to replicate the functionality of the external depth-sensing camera, providing an alternative pathway for a convenient implementation that would be more accessible. The next line of improvement involves the use of high dynamic range (HDR) and high-definition (HD) video capture. This would improve  support for capturing smartphones, smart watches, tablets, and digital screens in general by providing higher resolution and would reduce the  darkening of unlit elements in the scene, such as the users hands, providing a more realistic reproduction of  the user's skin tone, for instance. An alternative design that could improve screen readability in the VR would be to use a 2D-3D hybrid setup to track the smartphone and show users hands, following the approach of ~\cite{Bai2021}. This approach relies on a wired camera mounted on the HMD to track the device, but which could also eventually be replaced with access to the RGB stereo streams from the HMD built-in cameras, as suggested above.



\subsection{Implications of the findings}
 Despite the afore mentioned limitations, the results of the study provide several insights that inform VR interface design decisions related to this form of authentication. First, we have found that, while the current implementation of NRXR used in the study could soon be implemented in a more convenient way (with the removal of the external camera having the highest priority, as highlighted in Future Work), Near-Range Extended Reality can be effectively used to allow participants to access their smartphones to successfully achieve 2-factor authentication without the need for users to remove their HMD’s while in VR (RQ1). In accordance with the findings of the study, VR researchers and  developers may instead focus on developing alternatives to or variations of the checkers challenge (a novel technique), or continue relying on the use of the six-digit numeric PIN, as both options are  suitable and most preferred by users (RQ2 \&RQ4). Furthermore, VR  developers  can feel confident to avoid using  alphanumeric passwords, or limit their use to the first time users log into the system, as that was found to be the least suitable and least preferred option (RQ2 \& RQ4). Developers may also wish to use or support the use of secure password managers. Additionally, when faced with a design choice on whether users should use a gaze pointer or a VR controller for selecting items within the VR, the results suggest that the gaze pointer is the preferred option (RQ4). With respect to the question of whether the phone should be used as the device showing the challenge (Phone1\_* modalities), or the device used to respond to the challenge (HMD1\_Phone2 condition), the findings suggest that most people prefer to have the phone as the device showing the challenge, while the device for answering the challenge can be the HMD (RQ3). More generally, this work illustrates how NRXR could be used to mix the real and virtual worlds to facilitate certain tasks that require information from both environments and highlights the fact that the passthrough mode currently available in many HMD's can be refined in a way that  prioritizes access to the real-world elements found in close proximity of VR users, while allowing them to remain aware of the VR environment.

\section{Conclusions}
In this paper we explored methods for  implementing two-factor authentication in VR using smartphones. We demonstrate that it is possible for users to  perform such authentication successfully in many different ways. The results presented here suggest that NRXR-ID can be used to scan and select images to answer a CAPTCHA-style challenge, match visual patterns in a checkerboard-style, operate virtual numeric keypads, and read and type short passwords in VR for authentication purposes. 

We found that a checkers-style matching challenge is the most suitable option from among those considered, closely followed by the challenge of entering a six-digit numeric code using a virtual numeric keypad. This is significant, given that this option is the one most people are familiar with at this time, as the vast majority of digital service providers use two-factor authentication to provide access to their services. Other familiar methods, such as a CAPTCHA-style challenge, are also viable if the challenge is presented within the VR. We found that short but robust passwords are the most challenging to be entered and are also clearly disliked by users. There are many other variants to facilitate two-factor authentication in VR, and  further research is needed to explore these variants. 

This work shows that there are still open questions regarding the design and behaviour of the pass-through mode, which could be refined to  incorporate elements of the real world in a more selective approach to support the activities taking place inside the VR environment in a more targetted way. 



\appendix

\begin{figure*}[htbp]
        \centering
        \includegraphics[width=0.91\linewidth]{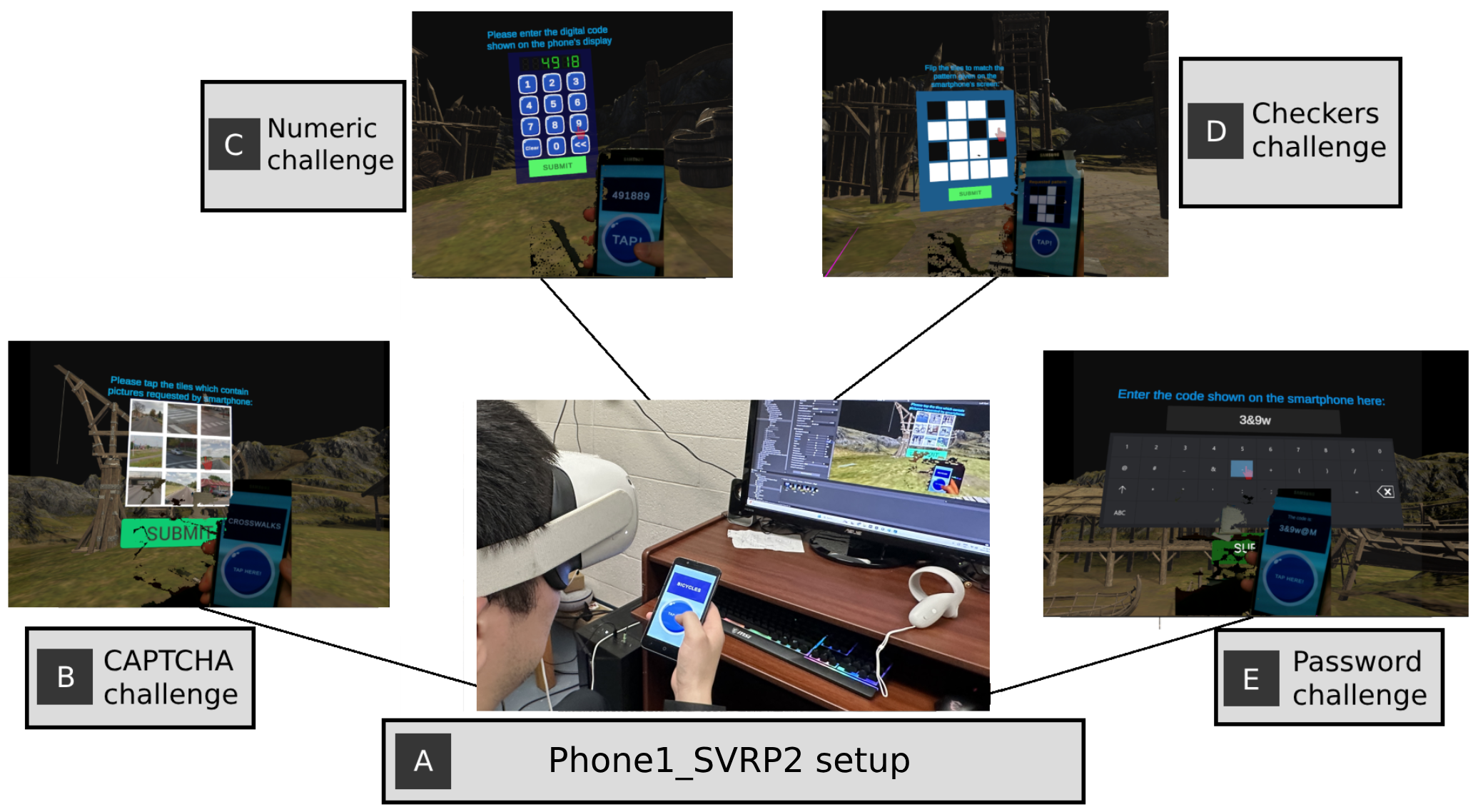}
        \caption{Overview of the authentication methods of condition 2 (Phone1\_SVRP2): A) Illustrates the experimental setup with a user holding the smartphone for reading and answering the challenge within the VR environment; B,C,D,E) Sequence of tasks provided for the user to solve, correspondingly: CAPTCHA challenge, numeric challenge, checkers challenge, password challenge.}
        \label{fig:challenges_condition2}
\end{figure*}

\begin{figure*}
    \centering
    \includegraphics[width=0.91\linewidth]{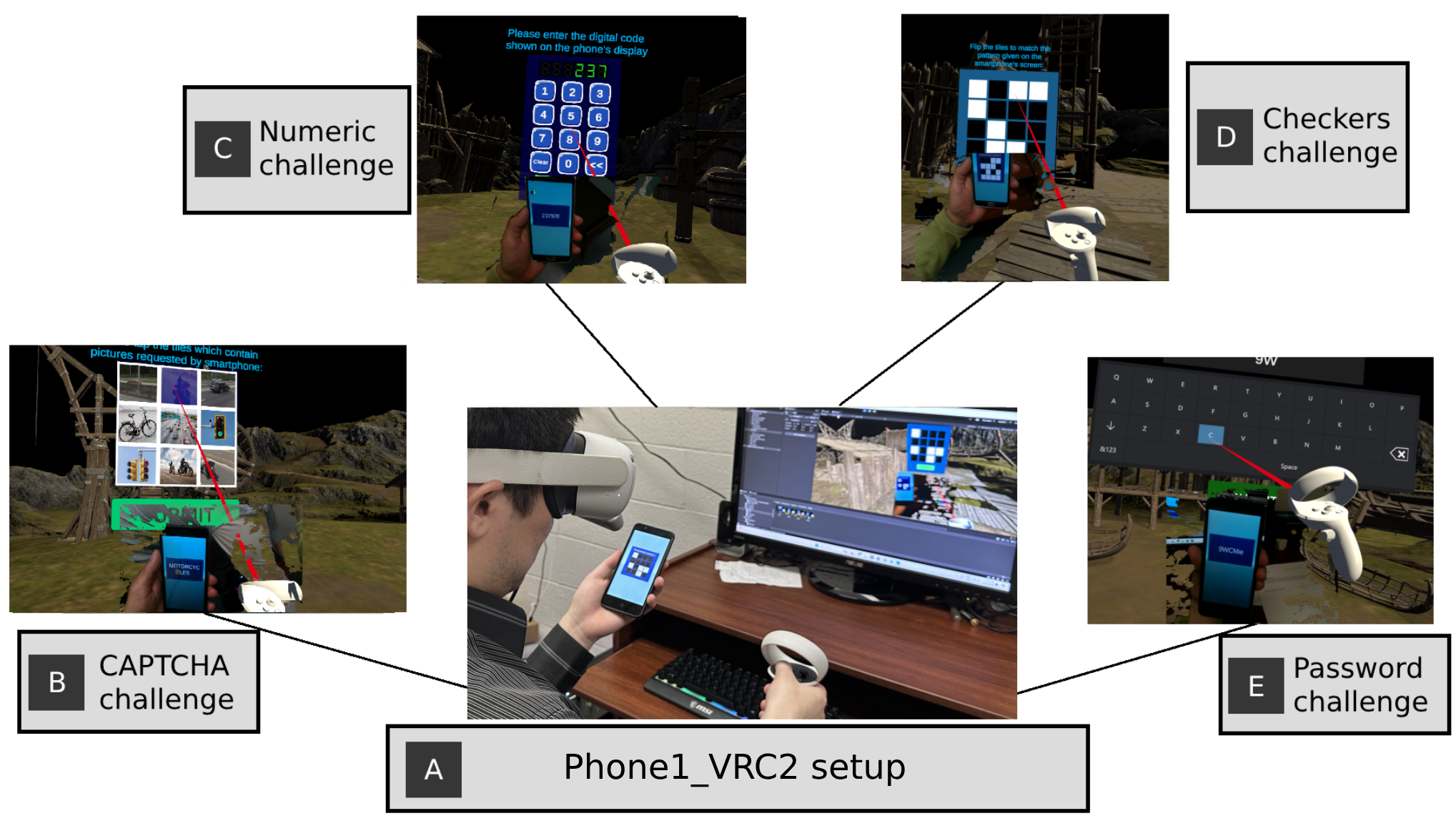}
        \caption{Overview of the authentication methods of condition 3 (Phone1\_VRC2): A) Illustrates the experimental setup with a user holding the smartphone and a VR controller; B,C,D,E) Sequence of tasks provided for the user to solve, correspondingly: CAPTCHA challenge, numeric challenge, checkers challenge, password challenge.}
        \label{fig:challenges_condition3}
\end{figure*}

    

\printbibliography

\end{document}